\documentclass{article}

\usepackage[preprint]{neurips_2026}

\usepackage[utf8]{inputenc}   
\usepackage[T1]{fontenc}      
\usepackage{hyperref}         
\usepackage{url}              
\usepackage{booktabs}         
\usepackage{amsfonts}         
\usepackage{amsmath}
\usepackage{amssymb}
\usepackage{nicefrac}         
\usepackage{microtype}        
\usepackage{xcolor}           
\usepackage{latexsym}

\usepackage{graphicx}
\usepackage{array}
\usepackage{multirow}
\usepackage{makecell}
\usepackage{enumitem}
\usepackage{longtable}
\usepackage{colortbl}
\usepackage{siunitx}
\usepackage{wrapfig}
\usepackage{needspace}
\usepackage{mdframed}
\usepackage{inconsolata}
\usepackage{tcolorbox}
\tcbuselibrary{skins, breakable}

\renewcommand{\cite}{\citep}
\setcitestyle{semicolon}

\AtBeginDocument{\setlength{\LTcapwidth}{\textwidth}}

\definecolor{lightblue}{rgb}{0.0, 0.5, 1.0}
\NewDocumentCommand{\minqian}{ mO{} }{\textcolor{lightblue}{\textsuperscript{\textit{Minqian}}\textsf{\textbf{\small[#1]}}}}

\NewDocumentCommand{\yuexi}{ mO{} }{\textcolor{teal}{\textsuperscript{\textit{Yuexi}}\textsf{\textbf{\small[#1]}}}}

\definecolor{revblue}{RGB}{0,70,170}

\newcommand{\revision}[1]{\textcolor{black}{#1}}

\newtcolorbox{promptbox}[1]{
    enhanced,
    breakable,
    title=#1,
    colback=gray!10,
    colframe=black!75,
    fonttitle=\bfseries,
    colbacktitle=gray!25,
    coltitle=black,
    arc=2mm,
    boxrule=0.5pt,
    drop shadow={black!50!white},
    left=5mm, right=5mm, top=3mm, bottom=3mm,
    parbox=true,
}

\title{Navigating Ideation Space:\\Decomposed Conceptual Representations for Positioning Scientific Ideas}

\author{%
  Minqian Liu\textsuperscript{1}\thanks{Minqian Liu and Yuexi Shen contributed equally.}\quad
  Yuexi Shen\textsuperscript{2}\footnotemark[1]\quad
  Dawei Zhou\textsuperscript{1}\quad
  Lifu Huang\textsuperscript{3} \\
  \textsuperscript{1}Virginia Tech\quad
  \textsuperscript{2}UC Santa Barbara\quad
  \textsuperscript{3}UC Davis
}

\begin{document}
\maketitle

\begin{abstract}

Scientific discovery requires new ideas to be situated within an ever-expanding landscape of existing knowledge. An emerging challenge is how to identify conceptually relevant prior work from rapidly growing literature, and assess how a new idea differentiates from existing research. 
Current embedding approaches typically conflate distinct conceptual aspects into single representations and cannot support fine-grained retrieval; meanwhile, LLM-based evaluators are subject to sycophancy biases, failing to provide discriminative novelty assessment. We introduce the \textbf{Ideation Space}, a structured representation that decomposes scientific knowledge into three distinct dimensions learned through contrastive training. Our framework enables principled measurement of conceptual distance between ideas and modeling of ideation transitions that capture the logical connections within an idea. Building upon this representation, we propose a \textbf{Hierarchical Sub-Space Retrieval} framework for efficient, targeted literature retrieval, and a \textbf{Decomposed Novelty Assessment} algorithm that identifies which aspects of an idea are novel. Extensive experiments demonstrate substantial improvements, achieving Recall@30 of 0.329 (16.7\% over the strongest baseline), and novelty assessment achieves promising correlation with expert judgments, outperforming all open-source LLM baselines up to 70B.
\end{abstract}

\section{Introduction}

Advancing scientific research \cite{wang2023scientific,aiscientist,wang2024scimon,si2025can} requires situating new ideas within the context of existing knowledge, which is increasingly difficult as the literature expands.
This pace has accelerated dramatically particularly in fields like computer science and AI. For instance, arXiv alone publishes over 20,000 papers monthly\footnote{\url{https://arxiv.org/stats/monthly_submissions}} in 2025 compared with around 6,000 monthly papers in 2010.
When developing a new research idea, a critical challenge emerges: \textit{how to find prior work that is conceptually relevant from massive literature, and determine how a new idea departs from what already exists}? 

Existing approaches to scientific literature retrieval and assessment suffer from three fundamental limitations. \textbf{First}, current embedding-based methods encode scientific papers into monolithic representations \cite{specter,singh2023scirepeval}. Such representations conflate distinct conceptual aspects, obscuring which aspect should drive the retrieval similarity.
\textbf{Second}, while recent work has turned to large language models (LLMs) for evaluating scientific hypotheses, LLM-based assessment often suffers from sycophancy biases \cite{sharma2024towards,ye2025justice}, producing scores that lack discriminative power. 
\textbf{Third}, existing methods do not evaluate \emph{which part} of an idea is conceptually novel, giving researchers little guidance on how to strengthen their work.

To address these challenges, we introduce \textbf{Ideation Space}, a structured representation that enables principled conceptual similarity measurement across decomposed dimensions of scientific contribution (\S\ref{sec:ideation_space}). 
Rather than representing papers as a single embedding, we construct three specialized sub-spaces, i.e., the \textit{problem} being addressed, the \textit{method} proposed, and the core \textit{findings} expected to discover, where each is learned through contrastive training to leverage the conceptual relationships in the citation network. We then utilize the three encoders to measure the \textit{conceptual distance} among scientific ideas. 
In addition, we propose to model the \textit{ideation transition} to capture logical connections within a proposed idea, such as how methodology is derived from the research problem.

Building on our ideation space, we propose a \textbf{Hierarchical Sub-Space Retrieval} framework that organizes literature into a structured database supporting efficient and targeted retrieval of conceptually relevant prior work (\S\ref{sec:retrieval}). Building upon retrieval, we introduce a graph-based \textbf{Decomposed Novelty Assessment}, where we measure novelty by quantifying the distance from retrieved evidence along each conceptual dimension (\S\ref{sec:novelty_eval}). Through extensive experiments, we demonstrate substantial improvements over existing methods. For retrieving conceptually similar literature, our approach achieves a Recall@30 of 0.329, outperforming the strongest baseline by 16.7\%. Our ideation transition method further enables retrieval of relevant reasoning patterns. In addition, our novelty assessment method shows a promising correlation of 0.37 with human judgments.
Our contributions are summarized as follows:
\begin{itemize}[leftmargin=*,nosep]
\item We introduce Ideation Space with decomposed representations of scientific knowledge through citation-contextualized contrastive learning, enabling the improved measurement of conceptual distance and ideation transition.
\item We propose a large-scale hierarchical sub-space retrieval framework that organizes scientific literature by conceptual function, which enables efficient and targeted retrieval. 
\item We develop a decomposed novelty evaluation pipeline that provides fine-grained assessments identifying where and why an idea is novel, achieving promising correlation with expert judgment comparable to frontier LLMs.
\end{itemize}



\section{Related Work}
\label{sec:related_work}


\paragraph{AI-Assisted Scientific Discovery and Novelty Assessment.}
Recent advances in LLMs have explored AI-assisted scientific discovery \cite{du-etal-2024-llms,aiscientist,kulkarni2025scientific,ghafarollahi2025sciagents,qi-etal-2025-metascientist} and its evaluation \cite{zheng2023judging,jin-etal-2024-agentreview,xeval,liu-etal-2024-holistic,feng2025grapheval,baek-etal-2025-researchagent}. 
For instance, \citet{si2025can} conducted the first large-scale study comparing LLM and human-expert ideas, while \citet{radensky2024scideator} developed Scideator for human-LLM collaborative ideation with retrieval-augmented novelty evaluation.
Recent works propose to evaluate scientific novelty by decomposing ideas \cite{qi-etal-2023-art,shahid-etal-2025-literature,afzal-etal-2026-beyond}. 
However, all three rely on LLMs at inference time for both decomposition and comparison, limiting scalability across large corpora. Our work departs in three ways: (1) we learn embedding sub-spaces via contrastive training on citation network, capturing conceptual relationships perceived by research communities; (2) our vector databases over 259K papers enable dimension-specific retrieval without LLM inference at query time; and (3) we uniquely model \emph{ideation transitions}, enabling retrieval of papers with analogous reasoning patterns, a capability absent from all concurrent works.

\paragraph{Scientific Knowledge Representations.}
Learning effective representations of scientific knowledge \cite{beltagy-etal-2019-scibert,zhang-etal-2023-content} is fundamental to literature retrieval 
in scientific landscapes \cite{wei2025ai}. 
\citet{specter} introduced SPECTER, training Transformer encoders using citation-informed contrastive learning. 
\citet{ostendorff-etal-2022-neighborhood} proposed SciNCL with neighborhood contrastive learning over citation graph embeddings, while \citet{singh2023scirepeval} developed SPECTER2 with task-specific adapters for multi-format representation. 
However, these methods encode papers into monolithic representations that conflate distinct conceptual aspects into a single vector, limiting their utility for fine-grained retrieval and novelty assessment. Our Ideation Space explicitly decomposes scientific knowledge into three functionally distinct sub-spaces, each learned through specialized contrastive training that leverages citation context.

\section{Ideation Space}
\label{sec:ideation_space}

We introduce \textbf{Ideation Space}, a structured framework for representing scientific knowledge that enables principled measurement of conceptual similarity across distinct dimensions of research contribution. Our key insight is that scientific ideas are not monolithic; instead, they comprise separable components serving different intellectual functions.

\begin{figure}[!tbp]
	\centering
	\includegraphics[width=\linewidth, trim={0 0 0 0},clip ]{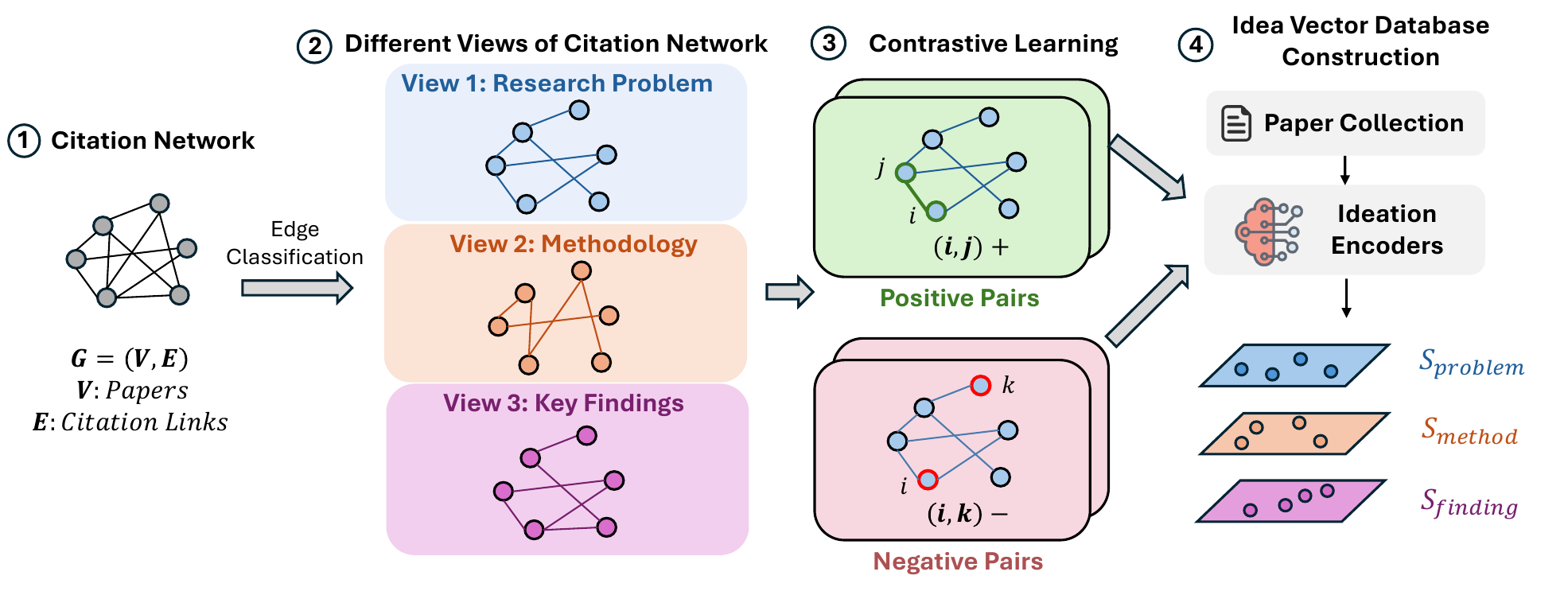}
	\caption{\textbf{Illustration of the Ideation Space construction process.} Starting from a citation network, we classify citation edges by relationships and decompose the graph into three complementary views. We then use these view-specific subgraphs to construct positive and negative pairs to train specialized encoders via contrastive learning, and produce three distinct vector databases for each sub-space.}
\label{fig:main_ideation}
\end{figure}

Our framework consists of three major steps. \textbf{First}, we partition scientific knowledge into three functionally distinct sub-spaces, i.e., \textit{problem}, \textit{method}, and \textit{findings}, each capturing a different aspect of contribution (\S\ref{sec:subspace_definition}). \textbf{Second}, we leverage the citation network to construct training data reflecting how researchers perceive conceptual relationships, using citation context to determine which dimension connects two papers (\S\ref{sec:training_data}). \textbf{Third}, we train specialized encoders via contrastive learning to produce embeddings where proximity reflects genuine conceptual similarity within each dimension (\S\ref{sec:contrastive_training}). The resulting ideation space supports both fine-grained similarity measurement and modeling of \textit{ideation transitions}, i.e., the logical connections within a proposed idea (\S\ref{sec:similarity_transition}). We show the overview of the ideation space construction process in Figure \ref{fig:main_ideation}.

\subsection{Sub-Space Decomposition}
\label{sec:subspace_definition}

We partition the collection of scientific literature $\mathcal{L}$ into three functionally distinct sub-spaces based on the conceptual function,
where $S_{\text{problem}}$ captures research problems and motivations, $S_{\text{method}}$ represents methodological approaches, and $S_{\text{findings}}$ indicates the expected results and conclusions. This decomposition can provide more comprehensive representations of scientific knowledge, such as a shared problem definition, a relevant methodology, or comparisons against reported findings. 
Each paper is embedded into these three spaces by specialized encoders $E_k$ ($k \in \{\text{problem}, \text{method}, \text{findings}\}$), enabling fine-grained conceptual distance measurement. 

Note that the problem-method-findings decomposition is structurally general and directly mirrors the IMRaD (Introduction, Methods, Results, and Discussion) format \cite{sollaci2004introduction}, which has been the dominant rhetorical structure across natural sciences, engineering, biomedicine, and social sciences for decades. Any empirical discipline that follows this organizational logic is, in principle, amenable to our framework.

\subsection{Training Data Construction}
\label{sec:training_data}

Building effective sub-space encoders requires training data reflecting genuine conceptual relationships within each dimension. We construct such data through a two-stage pipeline: identifying structurally related paper pairs from the citation graph, then classifying their functional relationships to assign them to appropriate sub-spaces. Our dataset comprises 259,340 papers from arXiv spanning 1993 to 2024, covering four major AI domains (cs.AI, cs.LG, cs.CV, and cs.CL), which ensures broad generalization.

\paragraph{Candidate Pair Identification.}
We extract candidate positive pairs based on structural proximity within a 5-year publication window. Two papers $(p, p^+)$ form a candidate pair if they have a direct citation link and satisfy at least one structural criterion: \textit{bibliographic coupling} (citing $\geq$10 common references) or \textit{co-citation} (co-cited $\geq$10 times). These thresholds ensure substantive intellectual connections, as papers meeting them typically engage with overlapping research questions or methodological traditions.

\paragraph{Function Classification.}
The key challenge is determining \textit{which} conceptual dimension connects a pair of related papers. We employ an LLM-based zero-shot classifier (i.e., Qwen3-4B-Instruct) that analyzes structured information about both papers (titles, research questions, methods, key findings) and produces confidence scores for three categories: \texttt{[Research Problem]}, \texttt{[Method Approach]}, and \texttt{[Key Findings]}. Pairs may be assigned to multiple sub-spaces when their relationship spans dimensions; those fitting no category are labeled \texttt{[Irrelevant]} and excluded. This distributes candidate pairs into three sub-graphs $G_{\text{problem}}$, $G_{\text{method}}$, and $G_{\text{findings}}$, each encoding dimension-specific conceptual relationships. 

\paragraph{Negative Sampling.}
Contrastive learning requires carefully constructed negatives. We employ two strategies. \textbf{In-batch negatives} sample from papers in the same batch that lack citation, co-citation, or bibliographic coupling relationships with the anchor, ensuring structural independence. \textbf{Hard conceptual negatives} exploit cross-dimensional relationships: we select pairs where the classifier assigns high scores to one category but low scores to another. For example, papers with high \texttt{[Research Problem]} but low \texttt{[Method Approach]} scores serve as hard negatives for $E_{\text{method}}$, forcing encoders to distinguish papers highly related in one dimension but less relevant in another.

\subsection{Contrastive Encoder Training}
\label{sec:contrastive_training}

We train three specialized encoders ($E_k$, $k \in \{\text{problem, method, findings}\}$) independently on their corresponding sub-graphs. Each encoder initializes from SPECTER2~\cite{singh2023scirepeval} to inherit general scientific knowledge, then fine-tunes to specialize in its designated dimension.

To ensure each encoder focuses on dimension-relevant content, we use an LLM to extract sub-space-specific text, i.e., research questions for $S_{\text{problem}}$, methodological descriptions for $S_{\text{method}}$, and key findings for $S_{\text{findings}}$. Each encoder trains exclusively on its corresponding text type.
For an anchor paper $p$ with positive set $P^+$ and negative set $P^-$, encoder $E_k$ minimizes:
\begin{equation}
\mathcal{L}_k = -\frac{1}{|P^+|} \sum_{p^+ \in P^+} \log \frac{e^{s(p, p^+)/\tau}}{\sum_{x \in P^+ \cup P^-} e^{\ell(p, x)}}
\end{equation}
where $s(\cdot, \cdot)$ denotes cosine similarity, $\tau$ is temperature, and:
\begin{equation}
\ell(p, x) = \begin{cases}
s(p, x)/\tau, & x \in P^+, \\
s(p, x)/\tau + \log w(x), & x \in P^-.
\end{cases}
\end{equation}
The weighting function $w(x) = ((s(p, x) + 1)/2)^\gamma$ emphasizes hard negatives by increasing their contribution to normalization. 

\subsection{Conceptual Similarity and Ideation Transition}
\label{sec:similarity_transition}

The trained encoders enable two complementary forms of analysis. For \textbf{conceptual similarity}, given a new idea $i$ and prior work $p$, we measure their similarity in sub-space $k$ as $\text{sim}_k(i, p) = \cos(E_k(i), E_k(p))$. Higher similarity indicates greater conceptual overlap along dimension $k$, suggesting reduced novelty in that aspect. 

Beyond static similarity, we capture the \textbf{ideation transition}, i.e., the reasoning structure within a paper, through transition vectors:
\begin{align}
T_{\text{PM}}(p) &= E_{\text{method}}(p) - E_{\text{problem}}(p) \\
T_{\text{MF}}(p) &= E_{\text{findings}}(p) - E_{\text{method}}(p)
\end{align}
The problem-to-method transition $T_{\text{PM}}$ captures how methodology addresses the research problem, while method-to-findings $T_{\text{MF}}$ represents how methodology leads to conclusions. These vectors enable the retrieval of papers with analogous reasoning patterns across different topics.

\section{Hierarchical Retrieval and Novelty Assessment}
\label{sec:applications}

Having established the ideation space as a structured representation of scientific knowledge, we now demonstrate its utility through two applications: retrieving conceptually relevant prior work and evaluating ideation novelty. These tasks are intrinsically interconnected: an idea's novelty is inversely related to how closely retrieved prior work resembles it. The more relevant the retrieved work, the less novel the idea. By leveraging the decomposed ideation space, we first retrieve prior work that is genuinely relevant along each conceptual dimension (\S\ref{sec:retrieval}), then assess novelty by measuring how much a new idea differentiates from the retrieved evidence (\S\ref{sec:novelty_eval}). This retrieval-grounded approach ensures that novelty scores reflect conceptual distance from the most relevant existing work, rather than surface-level textual similarity.


\subsection{Hierarchical Sub-Space Retrieval}
\label{sec:retrieval}

Traditional retrieval treats papers as atomic units, returning results based on overall topical similarity. However, researchers often seek prior work relevant to \textit{specific aspects}, e.g., papers addressing similar problems, employing related methods, or reporting comparable findings. Our framework addresses this need through structured databases supporting targeted, dimension-specific queries.

\paragraph{Database Construction.}
We construct five specialized databases from our corpus of 259,340 arXiv papers. Three \textbf{node databases} store sub-space-specific embeddings: for each paper, we extract dimension-specific text and encode it using the corresponding encoder ($E_{\text{problem}}$, $E_{\text{method}}$, $E_{\text{findings}}$). Two \textbf{transition databases} store the vectors $T_{\text{PM}}(p)$ and $T_{\text{MF}}(p)$ for each paper, enabling retrieval based on reasoning patterns.

\paragraph{Multi-Database Retrieval.}
Given a query paper, we perform parallel retrieval across all five databases using cosine similarity. Node retrieval identifies papers with similar problems, methods, or findings individually, while transition retrieval surfaces papers with analogous reasoning structures.
Together, these provide a comprehensive view of conceptually related prior work.


\subsection{Decomposed Novelty Assessment}
\label{sec:novelty_eval}

Beyond retrieval, we leverage ideation space for fine-grained novelty assessment. 
Our approach identifies \textit{which specific aspects} are novel and \textit{which} build upon existing work.

\paragraph{Reasoning Graph Extraction.}
We decompose the input idea $I$ into a reasoning graph $\mathcal{G}_I = (V_I, E_I)$ using an LLM with structured prompting. Unlike the simpler extraction for database construction, this produces a detailed graph representing the logical flow of the hypothesis. Nodes are classified into five types based on argumentative role: Background Context (BG) for established facts, Research Problem (RP) for limitations being addressed, Reasoning Insight (RI) for intermediate reasoning steps, Proposed Approach (PA) for concrete methodology, and Intended Contribution (CO) for expected outcomes. Edges represent logical dependencies, capturing argumentative flow from motivation through method to contribution.

\paragraph{Idea Encoding and Retrieval.}
We map each node to the appropriate sub-space based on its role: BG and RP nodes use $E_{\text{problem}}$; RI and PA nodes use $E_{\text{method}}$; CO nodes use $E_{\text{findings}}$. For each node $v \in V_I$, we query the corresponding database to retrieve similar prior claims, obtaining fine-grained evidence of which aspects have precedent.

\paragraph{Graph-Based Novelty Aggregation.}
We synthesize retrieval results using our novelty evaluation algorithm, which weights nodes by structural importance, where central arguments connecting multiple claims matter more than peripheral details. For each node $v$, we define $s(v)$ as the maximum similarity to any retrieved prior work. The importance weight combines two centrality measures:
\begin{equation}
w(v) = \frac{1}{2}\big(C_{\text{degree}}(v) + C_{\text{betweenness}}(v)\big)
\end{equation}
where $C_{\text{degree}}(v) = \frac{d(v)}{|V_I|-1}$ measures connectivity, and $C_{\text{betweenness}}(v) = \sum_{u \neq v \neq t} \frac{\sigma_{ut}(v)}{\sigma_{ut}}$ quantifies bridging importance ($\sigma_{ut}$ counts shortest paths from $u$ to $t$; $\sigma_{ut}(v)$ counts those through $v$). Weights are rescaled to $[0.5, 2.0]$ for balanced contribution.
The overall novelty score is:
\begin{equation}
N(I) = 1 - \frac{\sum_{v \in V_I} s(v) \cdot \tilde{w}(v)}{\sum_{v \in V_I} \tilde{w}(v)}
\end{equation}
This formulation ensures that novelty in structurally critical components, i.e., the intellectual core of the contribution, dominates the assessment. Beyond the aggregate score, for each node, we provide its similarity to the closest prior work, and the specific retrieved evidence, enabling researchers to understand not just \textit{how novel} their idea is, but \textit{what prior work} to engage with.

\section{Experiments}
\label{sec:experiments}

Our experiment targets two questions: Can our framework accurately retrieve the prior works relevant to a new idea's components? And does it generate novelty assessments that align with expert human judgment? We answer these through two complementary evaluations: \textbf{Hierarchical Retrieval Performance} and \textbf{Novelty Evaluation}.

\subsection{Experimental Setup}
\paragraph{Evaluation Datasets.}
For retrieval evaluation, we use 500 papers randomly selected from ICLR 2025 submissions as the gold test query set with human verification, and additional 1,330 papers as the silver test set that is fully automatically annotated. 
For novelty meta-evaluation, we use 93 research ideas from the AI-Researcher dataset~\cite{si2025can}, which includes expert-annotated novelty scores. As novelty judgment can be subjective, we include an inter-annotator agreement analysis in Appendix~\ref{app:iaa} to show the practical ceiling of the correlation.

\paragraph{Ground Truth Construction for Retrieval.}
We employ GPT-5.2 \cite{openai_gpt5_system_card} to curate ground truth by processing both references and OpenReview expert reviews.
\revision{To validate filtering quality, two CS/AI graduate students independently assessed a random sample of 30 query papers on a 3-point relevance scale, yielding an average score of \textbf{1.55/2.0}, confirming high reliability.
Our retrieval method is purely embedding-based at inference, so there is no circular dependency with the LLM-based ground truth construction.}
More details are shown in Appendix \ref{subsec:reference_extraction_prompts}.


\paragraph{Retrieval Baselines.}
We compare against strong sparse and dense retrieval models, including
(1) \textbf{BM25}~\cite{robertson2009probabilistic}, a traditional sparse keyword retrieval method;
(2) \textbf{E5$_{\text{base}}$-v2}~\cite{wang2022text}, a general-purpose semantic embedding model;
(3) \textbf{SPECTER2}~\cite{singh2023scirepeval}, a scientific document embedding model; and
(4) \textbf{SciNCL}~\cite{ostendorff-etal-2022-neighborhood}, a contrastive model optimized for citation graphs.
For node retrieval, each baseline constructs its own database using the concatenated title+abstract. For transition retrieval, we concatenate problem and approach texts to form transition representations.

\paragraph{Novelty Evaluation Baselines.}
For the novelty alignment task, we compare against two categories of baselines.
First is \textbf{LLM-as-a-Judge}, where
we prompt a set of large language models to assess novelty using the same evaluation rubric as expert annotators, including Qwen3-8B~\cite{yang2025qwen3}, Llama3.1-8B~\cite{dubey2024llama}, DeepSeek-R1-Distill-Llama-8B~\cite{guo2025deepseek}, and GPT-5 Mini and GPT-5 Nano~\cite{openai_gpt5_system_card}, proprietary models representing compact, production-grade language models.
The second is \textbf{Embedding-based Retrieval.}
We use E5$_{\text{base}}$-v2~\cite{wang2022text} embeddings with different input configurations (title only, summary only, and concatenated) to retrieve the top-1 most similar paper from the corpus, and compute novelty as $1 - \text{similarity}$, where lower similarity indicates higher novelty.


\paragraph{Implementation Details.}
We implement our framework using PyTorch and train all models on NVIDIA A40 GPUs. For the encoder, we initialize from SPECTER2~\cite{singh2023scirepeval} and fine-tune using the AdamW optimizer with a learning rate of $1 \times 10^{-5}$, employing a linear warmup for 700 steps followed by linear decay. We train for 10 epochs with a batch size of 32. For contrastive learning, we sample 2 positive and 8 negative examples per anchor, balancing memory efficiency and gradient stability. We use a temperature parameter of 0.1 for the InfoNCE loss and set the focal loss gamma to 2.0 to emphasize hard negatives. 

\subsection{Embedding Space Analysis}

\begin{wraptable}{r}{0.57\textwidth}
\vspace{-\intextsep}
\caption{Embedding space statistics on validation set: Positive/Negative means and margins across subspaces.}
\vspace{1mm}
\centering
\small
\setlength{\tabcolsep}{4pt}
\begin{tabular}{llccc}
\toprule
\textbf{Subspace} & \textbf{Model} & \textbf{Pos Mean} & \textbf{Neg Mean} & \textbf{Margin} \\
\midrule
\multirow{3}{*}{\makecell[l]{Research\\Problem}} 
 & Base       & 0.942 & 0.836 & 0.106 \\
 & Adapter    & 0.933 & 0.828 & 0.105 \\
 & \textbf{Fine-tuned} 
               & 0.792 & 0.058 & \textbf{0.735} \\
\midrule
\multirow{3}{*}{\makecell[l]{Method\\Approach}} 
 & Base       & 0.938 & 0.844 & 0.095 \\
 & Adapter    & 0.932 & 0.838 & 0.093 \\
 & \textbf{Fine-tuned} 
               & 0.654 & 0.010 & \textbf{0.644} \\
\midrule
\multirow{3}{*}{\makecell[l]{Key\\Findings}} 
 & Base       & 0.942 & 0.840 & 0.102 \\
 & Adapter    & 0.936 & 0.838 & 0.098 \\
 & \textbf{Fine-tuned} 
               & 0.697 & 0.020 & \textbf{0.678} \\
\bottomrule
\end{tabular}
\label{tab:embedding_stats}
\end{wraptable}

To understand the separation capabilities of our trained encoders, we analyze the positive and negative sample means and margins on the validation set across different subspaces. As shown in Table~\ref{tab:embedding_stats}, the fine-tuned models significantly increase the margin compared to base and adapter models, indicating superior discriminative power. The fine-tuned encoders achieve margins of 0.735, 0.644, and 0.678 for the three subspaces respectively, compared to ~0.1 for base models, demonstrating that our contrastive training effectively separates functionally relevant papers from irrelevant ones.

\begin{table}[t]
\centering
\caption{%
  Node retrieval performance (Recall, NDCG, Hit Rate) at $K \in \{9, 30, 90\}$.
  Baselines are evaluated using the concatenation of title and abstract as input.
}
\resizebox{\textwidth}{!}{%
\begin{tabular}{l *{3}{S[table-format=1.3]} *{3}{S[table-format=1.3]} *{3}{S[table-format=1.3]}}
\toprule
\multirow{2}{*}{\textbf{Model}}
  & \multicolumn{3}{c}{\textbf{Recall@K}}
  & \multicolumn{3}{c}{\textbf{NDCG@K}}
  & \multicolumn{3}{c}{\textbf{Hit Rate@K}} \\
\cmidrule(lr){2-4}\cmidrule(lr){5-7}\cmidrule(lr){8-10}
  & {$K{=}9$} & {$K{=}30$} & {$K{=}90$}
  & {$K{=}9$} & {$K{=}30$} & {$K{=}90$}
  & {$K{=}9$} & {$K{=}30$} & {$K{=}90$} \\
\midrule
BM25
  & 0.157 & 0.276 & 0.412 & 0.251 & 0.261 & 0.314 & 0.710 & 0.811 & 0.891 \\
E5$_{\textsc{base}}$-v2
  & 0.144 & 0.252 & 0.389 & 0.216 & 0.230 & 0.280 & 0.678 & 0.788 & 0.891 \\
SPECTER2
  & 0.160 & 0.282 & 0.423 & 0.242 & 0.257 & 0.310 & 0.667 & 0.811 & 0.893 \\
SciNCL
  & 0.157 & 0.281 & 0.427 & 0.233 & 0.250 & 0.305 & 0.697 & 0.788 & 0.886 \\
\midrule
\rowcolor{green!10} \textbf{Ideation Space} (Ours)
  & 0.169 & \textbf{0.329} & \textbf{0.483} & 0.250 & \textbf{0.277} & \textbf{0.335} & \textbf{0.702} & \textbf{0.861} & \textbf{0.927} \\
\quad w/ Subspace-RP
  & 0.160 & 0.300 & 0.450 & 0.238 & 0.264 & 0.322 & 0.678 & 0.831 & 0.897 \\
\quad w/ Subspace-MA
  & 0.167 & 0.299 & 0.453 & \textbf{0.251} & 0.270 & 0.328 & 0.697 & 0.835 & 0.901 \\
\quad w/ Subspace-KF
  & \textbf{0.172} & 0.304 & 0.444 & \textbf{0.251} & 0.270 & 0.324 & 0.695 & 0.820 & 0.886 \\
\bottomrule
\end{tabular}%
}
\label{tab:node_retrieval_results}
\end{table}

\subsection{Results of Hierarchical Retrieval}

We evaluate retrieval performance on two distinct tasks: \textbf{Node Retrieval} (identifying specific scientific entities, i.e., Problems, Methods, and Findings) and \textbf{Transition Retrieval} (identifying analogous reasoning patterns between these entities). We report Recall@K, NDCG@K, and Hit Rate@K to measure retrieval quality.

\paragraph{Node Retrieval Results.}
Table~\ref{tab:node_retrieval_results} presents the performance for retrieving scientific entities. Our Ideation Space consistently achieves the highest performance across all metrics. For instance, our Recall@90 reaches \textbf{0.483}, significantly outperforming the best baseline (SPECTER2 concat: 0.423). Even our individual subspaces (RP, MA, KF) often outperform the baselines, highlighting the effectiveness of decomposing scientific content into functional units rather than treating documents as monolithic blocks. The strong performance across different K values demonstrates that our approach retrieves relevant prior work at both high precision (low K) and high coverage (high K). We also show the results on the full baselines in Table \ref{tab:full_node_retrieval_results}, as well as the results on the silver test set in Table \ref{tab:expanded_retrieval_resultt}.

\paragraph{Transition Retrieval Results.}
Table~\ref{tab:transition_retrieval_results} presents the results for retrieving reasoning patterns. Identifying logical leaps is a more complex task than entity matching. While E5$_{\text{base}}$-v2 shows competitive ranking performance (NDCG), our \textbf{Transition Space} excels in \textbf{Recall@90 (0.275)} and \textbf{Hit Rate@90 (0.762)}. This demonstrates that explicitly modeling the vector transition between concepts ($T_{\text{PM}}$ and $T_{\text{MF}}$) is more effective at capturing diverse reasoning paths than static embeddings, providing a critical foundation for assessing methodological novelty.

\begin{wraptable}{r}{0.45\linewidth}
\centering
\footnotesize
\caption{Correlation between novelty estimation methods and ground-truth expert scores. \textbf{Bold}: best overall; \underline{underline}: best among non-proprietary methods.}
\vspace{1mm}
\begin{tabular}{lcc}
\toprule
\textbf{Method} & \textbf{Pearson} & \textbf{Spearman} \\
\midrule
Llama-3.1-8B & 0.053 & 0.089 \\
\quad + CoT & 0.034 & 0.052 \\
\arrayrulecolor{gray!50}\midrule\arrayrulecolor{black}
DS-R1-Distill-8B & 0.223 & 0.233 \\
\quad + CoT & 0.065 & 0.075 \\
\quad + Few-Shot & 0.296 & 0.241 \\
\arrayrulecolor{gray!50}\midrule\arrayrulecolor{black}
Qwen3-8B & 0.218 & 0.222 \\
\quad + Few-Shot & 0.282 & 0.293 \\
\arrayrulecolor{gray!50}\midrule\arrayrulecolor{black}
Qwen3-14B & 0.344 & 0.326 \\
\quad + Few-Shot & 0.323 & 0.323 \\
\arrayrulecolor{gray!50}\midrule\arrayrulecolor{black}
Llama-3-70B-It & 0.371 & 0.346 \\
Gemma3-27B & 0.336 & 0.309 \\
GPT-oss-20B & 0.195 & 0.198 \\
GPT-5-Mini & 0.369 & 0.385 \\
GPT-5.2 & \textbf{0.425} & \textbf{0.459} \\
\midrule
E5$_{\text{base}}$-v2 & 0.287 & 0.309 \\
SciNCL & 0.157 & 0.122 \\
SPECTER2 & 0.333 & 0.359 \\
\midrule
IdeationEval (Ours) & \underline{0.370} & \underline{0.384} \\
\bottomrule
\end{tabular}

\label{tab:novelty_corr_combined}
\end{wraptable}

\subsection{Results of Novelty Evaluation}
We evaluate whether our novelty assessments align with expert human judgment, measuring Pearson and Spearman correlation between predicted and ground-truth scores (Table~\ref{tab:novelty_corr_combined}).

\paragraph{Comparison with LLM-as-a-Judge.}
Our method outperforms all open-source LLM baselines regardless of scale, including Llama-3-70B-Instruct and Qwen3-14B with CoT. Among proprietary models, only GPT-5.2 substantially surpasses our approach, while GPT-5 Mini performs comparably. Notably, CoT degrades performance for DeepSeek-R1-Distill-Llama-8B and Llama-3.1-8B, suggesting unconstrained reasoning can introduce noise into novelty judgment; few-shot prompting is more reliable but still falls short of our method. Overall, our approach achieves competitive correlation \emph{without} LLM inference at evaluation time, offering substantially greater efficiency.

\paragraph{Comparison with Embedding-based Retrieval.}
Among embedding-based baselines, SPECTER2 performs best (Pearson: 0.333), followed by E5$_{\text{base}}$-v2 (0.287) and SciNCL (0.157). Our method substantially outperforms all three, demonstrating that decomposed sub-space representations with graph-theoretic importance weighting capture the multi-dimensional nature of novelty better than monolithic embedding similarity.

\subsection{Zero-Shot Cross-Domain Transfer}
\label{sec:cross_domain}
To test whether our framework depends on training data from the deployment domain, we evaluate our \emph{unmodified} CS-trained encoders on a biomedical retrieval test set of 472 bioRxiv queries, without any fine-tuning. Despite the substantial domain gap, our Ideation Space achieves the \textbf{best Hit Rate at every $K$} (HR@30 $= 0.976$) and matches the strongest baseline SPECTER2 on Recall ($0.740$ vs. $0.742$).
This suggests that our decomposition captures generalizable structure rather than domain-specific artifacts. Full results are in Appendix~\ref{app:cross_domain}.

\section{Conclusion}
We introduced the Ideation Space, a structured framework that decomposes scientific knowledge into three conceptually orthogonal dimensions, i.e., research problem, methodology, and core findings, enabling improved measurement of conceptual distance and ideation transitions. Building upon this representation, our Hierarchical Sub-Space Retrieval achieves up to 16.7\% improvement in recall, while our Decomposed Novelty Assessment attains 0.37 correlation with expert judgments. Our work establishes a promising paradigm for navigating the rapidly expanding scientific landscape, with potential applications in accelerating literature review, supporting peer review processes, and guiding researchers toward more impactful contributions.

\bibliographystyle{plainnat}
\bibliography{custom}


\appendix
\section{Limitations}

Our framework focuses on the AI and machine learning domain, with training data drawn from four arXiv categories (cs.AI, cs.LG, cs.CV, and cs.CL). While our zero-shot study on biomedical literature (Appendix~\ref{app:cross_domain}) shows competitive transfer to a neighboring field, systematically evaluating more distant disciplines such as physics or the social sciences remains an interesting direction for future work. Besides, our three-dimensional decomposition captures the core functional aspects of most scientific contributions, though certain interdisciplinary or theory-heavy works may exhibit more nuanced structures that could benefit from finer-grained representations. 

Additionally, our evaluation relies on citation relationships and LLM-assisted ground truth construction, which, while providing scalable and reproducible benchmarks, may not capture all dimensions of conceptual relevance that human experts perceive. The novelty meta-evaluation uses 93 research ideas from the AI-Researcher dataset, which, though expert-annotated, represents a specific subset of research styles. Moreover, our IAA analysis (Appendix~\ref{app:iaa}) shows that, expert agreement on novelty is itself only low-to-moderate, placing a practical ceiling on the model--expert correlation any automated method can achieve. Extending our evaluation to larger-scale human studies and diverse research communities would further validate the generalizability of our approach.

\begin{table}[t]
\centering
\vspace{4pt}
\caption{%
  Full node retrieval performance (Recall, NDCG, Hit Rate) at $K \in \{9, 30, 90\}$.
  Each sparse/dense baseline is evaluated on three input variants: title only,
  abstract only, and their concatenation (Concat).
  \textbf{Bold} denotes the best result per column.
}
\resizebox{\textwidth}{!}{%
\begin{tabular}{lc *{3}{S[table-format=1.3]} *{3}{S[table-format=1.3]} *{3}{S[table-format=1.3]}}
\toprule
\multirow{2}{*}{\textbf{Model}}
  & \multirow{2}{*}{\textbf{Input}}
  & \multicolumn{3}{c}{\textbf{Recall@K}}
  & \multicolumn{3}{c}{\textbf{NDCG@K}}
  & \multicolumn{3}{c}{\textbf{Hit Rate@K}} \\
\cmidrule(lr){3-5}\cmidrule(lr){6-8}\cmidrule(lr){9-11}
  & & {$K{=}9$} & {$K{=}30$} & {$K{=}90$}
    & {$K{=}9$} & {$K{=}30$} & {$K{=}90$}
    & {$K{=}9$} & {$K{=}30$} & {$K{=}90$} \\
\midrule
\multirow{3}{*}{BM25}
  & Title      & 0.087 & 0.159 & 0.232 & 0.140 & 0.152 & 0.180 & 0.481 & 0.622 & 0.719 \\
  & Abstract   & 0.143 & 0.246 & 0.365 & 0.230 & 0.236 & 0.281 & 0.674 & 0.794 & 0.856 \\
  & Concat & 0.157 & 0.276 & 0.412 & 0.251 & 0.261 & 0.314 & 0.710 & 0.811 & 0.891 \\
\arrayrulecolor{gray!50}\midrule\arrayrulecolor{black}
\multirow{3}{*}{E5$_{\textsc{base}}$-v2}
  & Title      & 0.110 & 0.180 & 0.264 & 0.169 & 0.175 & 0.207 & 0.573 & 0.682 & 0.779 \\
  & Abstract   & 0.134 & 0.239 & 0.372 & 0.204 & 0.218 & 0.268 & 0.635 & 0.773 & 0.873 \\
  & Concat & 0.144 & 0.252 & 0.389 & 0.216 & 0.230 & 0.280 & 0.678 & 0.788 & 0.891 \\
\arrayrulecolor{gray!50}\midrule\arrayrulecolor{black}
\multirow{3}{*}{SPECTER2}
  & Title      & 0.093 & 0.156 & 0.233 & 0.145 & 0.148 & 0.178 & 0.491 & 0.622 & 0.725 \\
  & Abstract   & 0.145 & 0.259 & 0.393 & 0.215 & 0.230 & 0.281 & 0.644 & 0.764 & 0.878 \\
  & Concat & 0.160 & 0.282 & 0.423 & 0.242 & 0.257 & 0.310 & 0.667 & 0.811 & 0.893 \\
\arrayrulecolor{gray!50}\midrule\arrayrulecolor{black}
\multirow{3}{*}{SciNCL}
  & Title      & 0.090 & 0.149 & 0.231 & 0.133 & 0.138 & 0.169 & 0.470 & 0.588 & 0.719 \\
  & Abstract   & 0.136 & 0.256 & 0.392 & 0.206 & 0.227 & 0.279 & 0.622 & 0.779 & 0.869 \\
  & Concat & 0.157 & 0.281 & 0.427 & 0.233 & 0.250 & 0.305 & 0.697 & 0.788 & 0.886 \\
\midrule
\rowcolor{green!10} \textbf{Ideation Space} (Ours) & RP+MA+KF
  & 0.169 & \textbf{0.329} & \textbf{0.483} & 0.250 & \textbf{0.277} & \textbf{0.335} & \textbf{0.702} & \textbf{0.861} & \textbf{0.927} \\
\quad w/ Subspace-RP & RP 
  & 0.160 & 0.300 & 0.450 & 0.238 & 0.264 & 0.322 & 0.678 & 0.831 & 0.897 \\
\quad w/ Subspace-MA & MA
  & 0.167 & 0.299 & 0.453 & \textbf{0.251} & 0.270 & 0.328 & 0.697 & 0.835 & 0.901 \\
\quad w/ Subspace-KF & KF
  & \textbf{0.172} & 0.304 & 0.444 & \textbf{0.251} & 0.270 & 0.324 & 0.695 & 0.820 & 0.886 \\

\bottomrule
\end{tabular}%
}

\label{tab:full_node_retrieval_results}
\end{table}

\begin{table}[t]
\centering
\vspace{4pt}
\caption{%
  Transition retrieval performance at $K \in \{9, 30, 90\}$. \
}
\resizebox{\textwidth}{!}{%
\begin{tabular}{l *{3}{S[table-format=1.3]} *{3}{S[table-format=1.3]} *{3}{S[table-format=1.3]}}
\toprule
\multirow{2}{*}{\textbf{Model}}
  & \multicolumn{3}{c}{\textbf{Recall@K}}
  & \multicolumn{3}{c}{\textbf{NDCG@K}}
  & \multicolumn{3}{c}{\textbf{Hit Rate@K}} \\
\cmidrule(lr){2-4}\cmidrule(lr){5-7}\cmidrule(lr){8-10}
  & {$K{=}9$} & {$K{=}30$} & {$K{=}90$}
  & {$K{=}9$} & {$K{=}30$} & {$K{=}90$}
  & {$K{=}9$} & {$K{=}30$} & {$K{=}90$} \\
\midrule
BM25
  & 0.076 & 0.135 & 0.229 & 0.140 & 0.135 & 0.169 & 0.432 & 0.593 & 0.739 \\
E5$_{\textsc{base}}$-v2
  & 0.095 & 0.160 & 0.247 & \textbf{0.168} & 0.160 & 0.192 & \textbf{0.492} & 0.627 & 0.730 \\
SPECTER2
  & 0.090 & 0.153 & 0.236 & 0.162 & 0.157 & 0.189 & 0.461 & 0.602 & 0.719 \\
SciNCL
  & 0.090 & 0.154 & 0.236 & 0.157 & 0.152 & 0.182 & 0.467 & 0.602 & 0.708 \\
\midrule
\rowcolor{green!10} \textbf{Transition Space} (Ours)
  & \textbf{0.099} & \textbf{0.180} & \textbf{0.275} & 0.159 & \textbf{0.166} & \textbf{0.203} & 0.465 & \textbf{0.643} & \textbf{0.762} \\
\quad w/ Subspace-P2M
  & 0.094 & 0.168 & 0.267 & 0.157 & 0.160 & 0.199 & 0.449 & 0.611 & 0.744 \\
\quad w/ Subspace-M2K
  & 0.078 & 0.148 & 0.237 & 0.131 & 0.135 & 0.169 & 0.393 & 0.548 & 0.710 \\
\bottomrule
\end{tabular}%
}
\label{tab:transition_retrieval_results}
\end{table}

\begin{table}[t]
\centering
\small
\caption{Node retrieval on the expanded test set with 1,330 papers annotated by Qwen3-4B. All baselines use concatenated title+abstract with $K{=}30$.}
\begin{tabular}{lccc}
\toprule
\textbf{Model} & \textbf{Recall@30} & \textbf{NDCG@30} & \textbf{Hit Rate@30} \\
\midrule
BM25 & 0.342 & 0.257 & 0.798 \\
E5$_\text{base}$-v2 & 0.322 & 0.236 & 0.788 \\
SPECTER2 & 0.345 & 0.247 & 0.804 \\
SciNCL & 0.337 & 0.240 & 0.796 \\
\midrule
\rowcolor{green!10} \textbf{Ideation Space} (Ours) & \textbf{0.388} & \textbf{0.271} & \textbf{0.820} \\
\quad w/ Subspace: RP & 0.360 & 0.257 & 0.801 \\
\quad w/ Subspace: MA & 0.367 & 0.266 & 0.802 \\
\quad w/ Subspace: KF & 0.365 & 0.262 & 0.813 \\
\bottomrule
\end{tabular}

\label{tab:expanded_retrieval_resultt}
\end{table}

\section{More Experiment Results}

\subsection{Qualitative Analysis: Novelty Evaluation Case Studies}
\label{sec:appendix_novelty_cases}

\begin{table}[t]
\centering
\caption{Case study of novelty evaluation.}
\resizebox{\textwidth}{!}{%
\begin{tabular}{lcccccccc}
\toprule
Paper Title & GT & Ideation Space (Ours) & DeepSeek & GPT-5-Nano & GPT-5-Mini & LLaMA-3.1-8B & Qwen3-8B & Qwen3-14B \\
\midrule
Adversarial Scenario Extrapolation & 5.50 & \textbf{5.53} & 9 & 7 & 6 & 8 & 8 & 8 \\
Probabilistic Opinion Pooling for QA & 5.67 & \textbf{9.79} & 8 & 7 & 6 & 8 & 8 & 8 \\
\bottomrule
\end{tabular}
}
\label{tab:novelty_case_study}
\end{table}

We provide detailed retrieved evidence for the two novelty evaluation cases discussed in \S\ref{sec:case_study}. For each case, Table~\ref{tab:novelty_detail} lists the top-1 retrieved paper per subspace along with its cosine similarity score.

\paragraph{Success Case: Adversarial Scenario Extrapolation.}
This proposal introduces a proactive defense framework in which LLMs generate and defend against hypothetical adversarial scenarios to improve robustness against novel attacks. As shown in the upper block of Table~\ref{tab:novelty_detail}, all three subspaces retrieve recent work (2023) directly addressing adversarial robustness for language models, with similarity scores consistently above 0.73. Specifically, the Research Problem retrieval identifies work on prompt-based adversarial robustness; the Method Approach retrieval surfaces a rewriting-based defense strategy closely related to the proposed generation-then-defense paradigm; and the Key Findings retrieval locates a safety classifier study whose evaluation scope aligns with the proposal's expected outcomes. These consistently high retrieval scores enable our graph-based algorithm to produce a well-calibrated novelty estimate of 5.53, nearly identical to the expert rating of 5.50.

\paragraph{Failure Case: Probabilistic Opinion Pooling for QA.}
This proposal applies probabilistic opinion pooling---a classical technique from statistics and decision theory---to handle conflicting evidence in question answering by combining weighted answer distributions from multiple sources. As the lower block of Table~\ref{tab:novelty_detail} shows, retrieval quality varies substantially across subspaces. While the Research Problem subspace retrieves a relevant paper on ambiguous QA (similarity $= 0.760$), the Method Approach subspace returns a 2015 paper on belief function combination with markedly lower similarity ($0.635$), failing to connect the proposal's core technique to its well-established roots. This terminology mismatch---``probabilistic opinion pooling inspired by social choice theory'' versus the classical ``linear opinion pooling''---constitutes the primary failure mode. The resulting reasoning graph contains an unusually high proportion of Research Insight nodes (8 out of 13), most with low retrieval scores, which our algorithm interprets as limited prior work and consequently inflates the novelty score to 9.79. Human experts, by contrast, recognize this as applying a known technique to a new domain and assign moderate novelty (5.67).

\begin{table}[t]
\centering
\small
\caption{Top-1 retrieved papers per subspace for the two novelty evaluation case studies. Similarity (Sim.) denotes cosine similarity in the corresponding ideation sub-space. GT: ground-truth expert novelty score.}
\resizebox{\textwidth}{!}{%
\begin{tabular}{llcc}
\toprule
\textbf{Subspace} & \textbf{Retrieved Paper} & \textbf{Sim.} & \textbf{Year} \\
\midrule
\multicolumn{4}{l}{\textit{Success: Adversarial Scenario Extrapolation} (GT = 5.50, Ours = 5.53)} \\
\midrule
Research Problem & Model-tuning Via Prompts Makes NLP Models Adversarially Robust & 0.817 & 2023 \\
Method Approach  & Don't Retrain, Just Rewrite: Countering Adversarial Perturbations by Rewriting Text & 0.735 & 2023 \\
Key Findings     & Robust Safety Classifier for Large Language Models: Adversarial Prompt Shield & 0.772 & 2023 \\
\midrule
\multicolumn{4}{l}{\textit{Failure: Probabilistic Opinion Pooling for QA} (GT = 5.67, Ours = 9.79)} \\
\midrule
Research Problem & Answering Ambiguous Questions through Generative Evidence Fusion and Round-Trip Prediction & 0.760 & 2020 \\
Method Approach  & Combining Partially Independent Belief Functions & 0.635 & 2015 \\
Key Findings     & Knowing More About Questions Can Help: Improving Calibration in Question Answering & 0.698 & 2021 \\
\bottomrule
\end{tabular}
}
\label{tab:novelty_detail}
\end{table}

\subsection{Comparison with Production Retrieval Systems}
\label{sec:human_eval}

\revision{To assess whether our method compares favorably against production retrieval systems, we conduct a human evaluation comparing our retrieval results against Semantic Scholar's paper search, a strong production system backed by a substantially larger index. Two annotators rated retrieved papers on a 3-point relevance scale (0 = not relevant, 1 = partially relevant, 2 = highly relevant), achieving substantial inter-annotator agreement (Cohen's $\kappa$ = 0.64). Our method obtains a mean relevance score of 1.59 versus Semantic Scholar's 1.02 (out of 2), demonstrating that our decomposed ideation space retrieves more conceptually relevant prior work even when compared to a widely-used production system backed by a far larger corpus. We note that these systems operate under fundamentally different conditions: they have access to the full web and proprietary indices, lack the temporal cutoff control necessary for fair novelty evaluation, and employ multi-step retrieval with substantial compute. In contrast, our framework is designed for efficient, reproducible, offline retrieval over a controlled corpus.}

\subsection{Cross-Domain Generalization Study}
\label{app:cross_domain}

A key concern with any specialized retrieval framework is whether its benefits depend on training data drawn from the exact domain of deployment. Our Ideation Space encoders are trained on AI/ML literature (cs.AI, cs.LG, cs.CV, cs.CL), which raises a natural question: \textit{does the decomposed problem-method-findings representation generalize to scientific disciplines with substantially different vocabulary, citation conventions, and conceptual structure?} To answer this, we evaluate our CS-trained encoders on biomedical literature without any fine-tuning, simulating a realistic deployment scenario in which practitioners apply pretrained scientific retrievers to a new field. This setting directly probes the \emph{domain transferability} of our framework and its \emph{robustness to distribution drift}.

\paragraph{Experimental Setup.}
We construct a biomedical test set following the same node retrieval protocol as our main experiments, with query papers and ground-truth relevant papers identified through citation relationships within the biomedical literature.
The corpus consists of \num{63082} papers drawn from bioRxiv (2013--2026), from which we sample 500 query papers published in 2026; after filtering queries with no retrievable ground-truth, 472 queries remain.
We compare our Ideation Space against four strong baselines: BM25, E5$_{\text{base}}$-v2, SPECTER2, and SciNCL.
Each baseline is applied directly off the shelf; for brevity, we report only the concat (title + abstract) variant, which consistently outperforms the title-only and abstract-only variants.
Crucially, our Ideation Space encoders are \emph{not} retrained on biomedical data---we apply the exact same checkpoints used in the main experiments.
We report Recall@$K$ and Hit Rate@$K$ at $K \in \{3, 10, 30\}$.

\begin{table}[ht]
\centering
\small
\caption{\textbf{Zero-shot cross-domain transfer on biomedical domain.} Our Ideation Space encoders are trained exclusively on AI/ML papers and applied to biomedical retrieval queries without any fine-tuning. Baselines are off-the-shelf retrievers; only the concat (title + abstract) variant is reported. The best result per column is highlighted in bold.}
\setlength{\tabcolsep}{4.5pt}
\renewcommand{\arraystretch}{1.08}
\begin{tabular}{l|ccc|ccc}
\toprule
\multirow{2}{*}{Model} &
\multicolumn{3}{c|}{Recall @ K} &
\multicolumn{3}{c}{Hit Rate @ K} \\
& 3 & 10 & 30 & 3 & 10 & 30 \\
\midrule
BM25 
& 0.392 & 0.592 & 0.729
& 0.905 & 0.939 & 0.962 \\
E5$_{\text{base}}$-v2 
& 0.391 & 0.587 & 0.723
& 0.898 & 0.949 & 0.970 \\
SPECTER2 
& \textbf{0.393} & \textbf{0.609} & \textbf{0.742}
& 0.896 & 0.955 & 0.972 \\
SciNCL 
& 0.362 & 0.557 & 0.708
& 0.875 & 0.934 & 0.964 \\
\midrule
\textbf{Ideation Space}
& 0.373 & 0.586 & 0.740
& \textbf{0.904} & \textbf{0.961} & \textbf{0.976} \\
Subspace-RP
& 0.358 & 0.557 & 0.712
& 0.879 & 0.944 & 0.974 \\
Subspace-MA
& 0.372 & 0.566 & 0.722
& 0.891 & 0.957 & \textbf{0.976} \\
Subspace-KF
& 0.358 & 0.569 & 0.714
& 0.885 & 0.942 & 0.964 \\

\bottomrule
\end{tabular}
\label{tab:retrieval_bio_concat}
\end{table}

\paragraph{Results.} Table~\ref{tab:retrieval_bio_concat} reports zero-shot transfer performance. Despite the substantial domain gap between AI/ML and biomedical literature, our Ideation Space remains highly competitive and achieves the best Hit Rate across all three $K$ values (HR@3 = 0.904, tied with BM25; HR@10 = 0.961; HR@30 = 0.976). On Recall, our method (R@30 = 0.740) is on par with the strongest baseline SPECTER2 (R@30 = 0.742) and outperforms BM25, E5$_{\text{base}}$-v2, and SciNCL by clear margins. We highlight three key observations:

\noindent\textbf{(i) Robustness to distribution drift.} Although SPECTER2 holds a slight edge on Recall, this is expected: SPECTER2 is a general-purpose scientific encoder pretrained on broad scientific corpora that already include biomedical text, whereas our encoders are specialized to AI/ML citation patterns. The fact that our \emph{domain-specialized} encoders match a \emph{domain-general} baseline under distribution shift is itself a strong signal of transferability, indicating that our framework does not overfit to the surface features of AI/ML literature.

\noindent\textbf{(ii) Consistent Hit Rate gains.} Our method outperforms all baselines on Hit Rate at every $K$, with a small but consistent edge over SPECTER2 at $K{=}30$ (0.972 vs.\ 0.976). Hit Rate measures whether at least one relevant document appears in the top-$K$, which is particularly important for downstream novelty assessment: missing relevant prior work can lead to overestimated novelty scores. The consistent Hit Rate advantage suggests that the decomposed sub-spaces continue to surface diverse, complementary evidence even in an unfamiliar domain.

\noindent\textbf{(iii) Sub-space ablations remain meaningful.} The three single-aspect variants (RP, MA, KF) each retain distinct retrieval characteristics under transfer, with Subspace-MA achieving the best HR@30 among ablations. This indicates that the functional decomposition learned from AI/ML citation contexts captures generalizable structure rather than domain-specific artifacts.

\paragraph{Discussion.} These results support the claim that the Ideation Space framework is \emph{domain-transferable by design}. From a practical standpoint, our released CS-trained encoders can be directly applied to neighboring scientific domains without expensive domain-specific retraining, lowering the barrier for adoption in fields where large citation-annotated corpora may not be readily available. We leave the systematic study of more distant domains (e.g., physics, social sciences) and the analysis of when domain-specific retraining becomes necessary as directions for future work.

\subsection{Inter-Annotator Agreement on Expert Novelty Scores}
\label{app:iaa}

\begin{wraptable}{r}{0.46\textwidth}
\vspace{-\intextsep}
\caption{Inter-annotator agreement on expert novelty scores from the AI-Researcher dataset~\cite{si2025can}, computed over all 98 ideas. $\alpha$ is reported under both ordinal and interval distance metrics. $|\text{diff}|$ is the mean absolute pairwise difference between annotators on the 1--10 scale. \textbf{Exact} and \textbf{Within-1} report the proportion of annotator pairs whose scores match exactly or fall within one point.}
\vspace{2mm}
\centering
\small
\begin{tabular}{lc}
\toprule
\textbf{Metric} & \textbf{Value} \\
\midrule
N (ideas) & 98 \\
Krippendorff's $\alpha$ (ordinal) & 0.236 \\
Krippendorff's $\alpha$ (interval) & 0.258 \\
Mean $|\text{diff}|$ & 1.80 \\
Exact agreement & 19.6\% \\
Within-1 agreement & 49.1\% \\
Grand mean score & 5.24 \\
\bottomrule
\end{tabular}
\label{tab:iaa}
\end{wraptable}

A natural question for our novelty meta-evaluation is what level of model--expert correlation should be considered strong, given that novelty is an inherently subjective judgment. To contextualize the correlations reported in \S\ref{sec:novelty_eval}, we conduct an inter-annotator agreement (IAA) analysis on the AI-Researcher dataset~\cite{si2025can}, which provides multiple independent expert ratings per idea on a 1--10 novelty scale. Table~\ref{tab:iaa} summarizes the agreement statistics across all 98 ideas.

We report Krippendorff's $\alpha$ under both \textbf{ordinal} and \textbf{interval} distance metrics, along with three complementary measures: the mean absolute pairwise difference between annotators ($|\text{diff}|$), exact agreement rate, and within-1-point agreement rate. We observe $\alpha = 0.236$ (ordinal) and $0.258$ (interval), with a mean pairwise disagreement of $1.80$ points and only $19.6\%$ exact agreement, though $49.1\%$ of pairs fall within 1 point.

These low-to-moderate agreement values reveal that \textbf{novelty assessment is intrinsically a high-variance task even among domain experts}, consistent with prior peer-review studies. This finding establishes a meaningful ceiling on achievable model--expert correlation: an automated method cannot be expected to agree with any single expert more strongly than experts agree with one another. Viewed against this ceiling, our Ideation Space achieves Pearson $r = 0.370$ and Spearman $\rho = 0.384$ with averaged expert scores (\S\ref{sec:novelty_eval}), which is on par with the level of agreement between expert pairs themselves. This reframes our result: rather than being a modest absolute correlation, $0.37$ approaches the practical upper bound imposed by human disagreement, and is achieved entirely without LLM inference at evaluation time.

\begin{table}[ht]
\centering
\small
\caption{Comparison between jointly trained retrieval and Ideation Space retrieval. Bold denotes the best result per column.}
\setlength{\tabcolsep}{3.5pt}
\renewcommand{\arraystretch}{1.05}

\begin{tabular}{l|ccc|ccc|ccc}
\toprule
Model
& \multicolumn{3}{c|}{Recall @ K}
& \multicolumn{3}{c|}{NDCG @ K}
& \multicolumn{3}{c}{Hit Rate @ K} \\
& 9 & 30 & 90
& 9 & 30 & 90
& 9 & 30 & 90 \\
\midrule

Jointly Trained
& \textbf{0.172} & 0.304 & 0.444
& \textbf{0.251} & 0.270 & 0.324
& 0.695 & 0.820 & 0.886 \\

Ideation Space
& 0.169 & \textbf{0.329} & \textbf{0.483}
& 0.250 & \textbf{0.277} & \textbf{0.335}
& \textbf{0.702} & \textbf{0.861} & \textbf{0.927} \\

\bottomrule
\end{tabular}
\label{tab:joint_vs_ideation}

\end{table}

\section{More Experimental Details}
\label{sec:more_exp_details}

\subsection{More Details of Structured Extraction}

We use Qwen3-8B (\texttt{Qwen/Qwen3-8B}) to extract structured representations from paper abstracts. The model decomposes each abstract into three atomic dimensions---Research Problem, Method Approach, and Key Findings---using the following configuration: temperature $= 0.0$ for deterministic outputs, max\_tokens $= 512$ to accommodate detailed three-sentence responses, top-$p = 0.95$ for controlled sampling, and repetition\_penalty $= 1.1$ to discourage redundant phrasing.

\subsection{More Details of Building Ground Truth for Node Retrieval}
\label{subsec:reference_extraction_prompts}

\revision{For retrieval evaluation, we curate ground truth by processing both references and OpenReview expert reviews. Specifically, we extract all references cited by each test paper and use an LLM to classify their relevance for novelty assessment. The LLM receives structured information about both papers (title, research question, method, key findings) and determines whether the cited work is pertinent to evaluating novelty. In addition, we extract papers mentioned in human reviews as relevant to the novelty assessment. Our filtering process identifies papers related along three dimensions as well as papers that share analogous reasoning patterns between adjacent nodes.}

\revision{To validate the LLM filtering quality, we conduct a human validation study. We randomly sample 30 query papers from the test set, and two CS/AI graduate students independently verify whether the LLM-filtered ground-truth papers are genuinely relevant using a 3-point scale (0 = not relevant, 1 = partially relevant, 2 = highly relevant). The resulting average score is \textbf{1.55} (out of 2), confirming that the vast majority of filtered papers are highly relevant and that the LLM filter does not introduce meaningful bias.}

For ground-truth extraction from human reviews, we use GPT-5.2~\cite{openai_gpt5_system_card} (via the OpenAI Responses API) to extract novelty-related references from reviewer comments. The model analyzes reviewer feedback and categorizes cited works into three dimensions: Research Problem Overlap, Method/Approach Overlap, and Key Findings Overlap. We use the following configuration: max\_output\_tokens $= 2048$ to accommodate detailed structured output, reasoning effort set to \texttt{`medium'}, and text verbosity set to \texttt{`medium'}.

\section{Ablation Study}
\label{sec:ablation}

A central design choice in our framework is the decomposition of paper representations into three subspaces---\textit{research problem}, \textit{method approach}, and \textit{key findings}---with a dedicated retriever trained for each. To assess whether this decomposition is necessary, we compare against a jointly trained variant that merges all subspace supervision into a single retriever.

\paragraph{Setup.}
For the jointly trained variant, we take the union (deduplicated by paper ID) of positives and hard negatives across the three subspaces, and encode each paper as the concatenation of its \texttt{research\_question}, \texttt{method\_approach}, and \texttt{key\_findings} fields joined by \texttt{[SEP]}. All other hyperparameters match the per-subspace setup. At inference time, the joint model returns a single ranked list, while Ideation Space merges the top results from three independent retrievers.

\paragraph{Results.}
Table~\ref{tab:joint_vs_ideation} reports Recall, NDCG, and Hit Rate at $K \in \{9, 30, 90\}$. At $K = 9$, the two variants are essentially tied, with the joint model slightly ahead on Recall and NDCG. However, as $K$ grows, Ideation Space pulls ahead on every metric: at $K = 90$, it improves Recall by $8.8\%$, NDCG by $3.4\%$, and Hit Rate by $4.6\%$ relative to the joint model.

We attribute this gap to two complementary factors. First, each subspace retriever is trained on a single, homogeneous notion of relevance, which sharpens its ranking signal. Second, because the three retrievers attend to different semantic axes, their top-$K$ lists are only partially overlapping, so their union covers more facets of relevance than any single model can. The joint model, by collapsing three views into one embedding, favors conservative neighbors that score moderately well across all axes, leading to a flatter recall curve. This breadth-of-coverage advantage is precisely what downstream ideation tasks require.

\section{Qualitative Analysis: Retrieval Case Studies}
\label{sec:appendix_case_studies}

To provide a concrete illustration of how multi-dimensional retrieval yields more targeted evidence, we present a detailed comparison for the query paper \textit{LongHalQA}~(Table~\ref{tab:case_study_detail}). For each ideation dimension, we show the top-1 retrieved paper's title, relevant text, and similarity score, alongside the top-5 results from the strongest baseline (SPECTER2, title + abstract).

{\small
\begin{longtable}{>{\raggedright\arraybackslash}m{2.2cm} m{\dimexpr\textwidth-2.2cm-1.1cm-6\tabcolsep\relax} >{\centering\arraybackslash}m{1.1cm}}
\caption{Detailed retrieval comparison for \textit{LongHalQA}. The Query Paper details are presented first. For each dimension, we display the top retrieved paper's Title and relevant Text. Our retrieval surfaces specific hallucination benchmarks (AMBER, HaELM) while the baseline focuses on general MLLM benchmarks.}
\label{tab:case_study_detail} \\
\toprule
\multicolumn{3}{l}{\textbf{Query Paper Details}} \\
\midrule
\multicolumn{3}{m{\dimexpr\textwidth-2\tabcolsep\relax}}{\textbf{Title:} LongHalQA: Long-Context Hallucination Evaluation for MultiModal Large Language Models} \\
\addlinespace[0.5em]
\multicolumn{3}{m{\dimexpr\textwidth-2\tabcolsep\relax}}{\textbf{Research Problem:} This work addresses the gap in existing benchmarks for evaluating hallucinations in multimodal large language models (MLLMs), which rely on \textbf{simplistic discriminative questions} or \textbf{computationally intensive LLM evaluators}.} \\
\addlinespace[0.5em]
\multicolumn{3}{m{\dimexpr\textwidth-2\tabcolsep\relax}}{\textbf{Method Approach:} The paper introduces LongHalQA, an \textbf{LLM-free benchmark} featuring \textbf{6K long and complex hallucination texts} generated by GPT4V, along with two unified tasks---hallucination discrimination and completion.} \\
\addlinespace[0.5em]
\multicolumn{3}{m{\dimexpr\textwidth-2\tabcolsep\relax}}{\textbf{Key Findings:} Extensive experiments on recent MLLMs reveal \textbf{significant challenges in handling long and complex hallucination texts}, highlighting the necessity for advanced benchmarks like LongHalQA.} \\
\midrule
\multicolumn{1}{c}{\textbf{Dimension}} & \multicolumn{1}{c}{\textbf{Retrieved Paper (Title \& Text)}} & \multicolumn{1}{c}{\textbf{Score}} \\
\midrule
\endfirsthead

\caption[]{(continued)} \\
\toprule
\multicolumn{3}{c}{\tablename\ \thetable\ -- \textit{Continued from previous page}} \\
\midrule
\multicolumn{3}{m{\dimexpr\textwidth-2\tabcolsep\relax}}{\textbf{Query Paper:} \textit{LongHalQA: Long-Context Hallucination Evaluation for MultiModal Large Language Models}} \\
\midrule
\multicolumn{1}{c}{\textbf{Dimension}} & \multicolumn{1}{c}{\textbf{Retrieved Paper (Title \& Text)}} & \multicolumn{1}{c}{\textbf{Score}} \\
\midrule
\endhead

\midrule
\multicolumn{3}{r}{\textit{Continued on next page...}} \\
\endfoot

\bottomrule
\endlastfoot


\multicolumn{3}{l}{\textit{\textbf{Our Multi-Dimensional Retrieval (Top 1 per Dimension)}}} \\
\midrule

Research Problem & 
\textbf{Title:} SEED-Bench-2: Benchmarking Multimodal Large Language Models 
\newline 
\textbf{Text:} This work addresses the gap in existing MLLM benchmarks by identifying the need for a \textbf{comprehensive evaluation framework} that assesses the \textbf{hierarchical multimodal capabilities} of models beyond single-image-text comprehension. 
& 0.893 \\
\midrule

Method Approach & 
\textbf{Title:} An LLM-free Multi-dimensional Benchmark for MLLMs Hallucination Evaluation 
\newline 
\textbf{Text:} The paper introduces AMBER, an \textbf{LLM-free, multi-dimensional benchmark} that efficiently evaluates \textbf{hallucination types} across a wide range of tasks, using a \textbf{low-cost and scalable evaluation pipeline}. 
& 0.826 \\
\midrule

Key Findings & 
\textbf{Title:} Evaluation and Analysis of Hallucination in Large Vision-Language Models 
\newline 
\textbf{Text:} HaELM achieves approximately 95\% performance comparable to ChatGPT, demonstrating strong effectiveness in \textbf{evaluating hallucination} while providing practical benefits such as \textbf{local deployment}. 
& 0.835 \\
\midrule

Problem $\to$ Method & 
\textbf{Title:} Hallusionbench: An Advanced Diagnostic Suite for Entangled Language Hallucination and Visual Illusion 
\newline 
\textbf{Text:} This work addresses the unique challenge of evaluating image-context reasoning in large visual-language models (LVLMs) by focusing on the \textbf{nuanced detection of hallucination and visual illusion}. 
& 0.660 \\
\midrule

Method $\to$ Findings & 
\textbf{Title:} An LLM-free Multi-dimensional Benchmark for MLLMs Hallucination Evaluation 
\newline 
\textbf{Text:} AMBER demonstrates significant practical value by providing a comprehensive assessment of mainstream MLLMs like GPT-4V, revealing \textbf{performance disparities} and offering \textbf{actionable guidelines to mitigate hallucinations}. 
& 0.682 \\
\midrule

\multicolumn{3}{l}{\textit{\textbf{Baseline Title \& Abstract Based Retrieval (Top 5)}}} \\
\midrule

Rank 1 & 
\textbf{Title:} SEED-Bench-2: Benchmarking Multimodal Large Language Models 
\newline 
\textbf{Text:} This work addresses the gap in existing MLLM benchmarks by identifying the need for a \textbf{comprehensive evaluation framework} that assesses the \textbf{hierarchical multimodal capabilities} of models. 
& 0.893 \\
\midrule

Rank 2 & 
\textbf{Title:} MME: A Comprehensive Evaluation Benchmark for Multimodal Large Language Models 
\newline 
\textbf{Text:} This work addresses the gap in comprehensive evaluation of Multimodal Large Language Models (MLLMs), as existing case studies fail to provide a \textbf{holistic assessment of their perception and cognition capabilities}. 
& 0.879 \\
\midrule

Rank 3 & 
\textbf{Title:} MLLM-Bench: Evaluating Multimodal LLMs with Per-sample Criteria 
\newline 
\textbf{Text:} This work addresses the gap in evaluating multimodal large language models (MLLMs) by focusing on the challenge of assessing creative and associative tasks through \textbf{user experience-driven metrics}. 
& 0.877 \\
\midrule

Rank 4 & 
\textbf{Title:} SEED-Bench: Benchmarking Multimodal LLMs with Generative Comprehension 
\newline 
\textbf{Text:} This work addresses the lack of comprehensive evaluation frameworks for generative comprehension in multimodal large language models (MLLMs), focusing on the unique challenge of assessing both \textbf{image and video understanding}. 
& 0.626 \\
\midrule

Rank 5 & 
\textbf{Title:} TinyLVLM-eHub: Towards Comprehensive and Efficient Evaluation for Large Vision-Language Models 
\newline 
\textbf{Text:} This work addresses the gap in evaluating Large Vision-Language Models (LVLMs) by focusing on a comprehensive assessment of their multimodal capabilities, particularly in areas such as \textbf{object hallucination}. 
& 0.625 \\

\end{longtable}
}

\section{LLM-Based Annotation and Prompt Configuration}

\label{sec:prompt_configs}

\subsection{Structured Extraction from Paper Abstracts}
We use Qwen3-8B (model path: \texttt{Qwen/Qwen3-8B}) to extract structured representations from paper abstracts. The model decomposes each abstract into three atomic dimensions: Research Problem, Method Approach, and Key Findings, with the following configuration: temperature=0.0 for deterministic outputs, max\_tokens=512 to accommodate detailed three-sentence responses, top-p=0.95 for controlled randomness, and repetition\_penalty=1.1 to prevent redundant phrasing. Below is the complete prompt used for extraction:

\begin{tcolorbox}[colback=gray!5!white, colframe=gray!75!black, breakable]
\small
\texttt{You are an expert academic writer and editor tasked with creating precise, informative structured summaries of research paper abstracts. Your goal is to distill the essence of complex research into a clear and concise summary that captures the paper's unique contributions and distinguishes it from related work.}

\vspace{0.3cm}
\textbf{Task Instructions:}

\texttt{Transform the following academic paper abstract into exactly THREE sentences. Each sentence should address one of the required aspects and be self-contained and clear to understand for researchers in related fields. Each sentence should be both concise and informative, maintaining the key information while being brief and not verbose.}

\vspace{0.3cm}
\textbf{Required Elements to Include (One sentence each):}

\begin{enumerate}[leftmargin=*,itemsep=0pt]
    \item \textbf{Research Question/Problem}: What specific challenge, gap, or question does this work uniquely address that distinguishes it from existing research?
    \begin{itemize}[itemsep=0pt]
        \item Focus on the novel problem formulation or unique angle the paper takes
        \item Identify what specific gap in current literature this work fills
        \item Emphasize the distinctive aspect of the research question that sets it apart
    \end{itemize}
    
    \item \textbf{Method/Approach}: What novel technique, algorithm, methodology, or experimental design is introduced that sets this work apart from previous approaches?
    \begin{itemize}[itemsep=0pt]
        \item Highlight the innovative technical contribution or methodological advancement
        \item Describe the unique algorithmic approach, architecture, or experimental setup
        \item Emphasize what makes the method distinct from existing solutions
    \end{itemize}
    
    \item \textbf{Key Findings/Insights/Contributions}: What are the main empirical results, theoretical insights, or practical contributions that represent the paper's unique value and impact?
    \begin{itemize}[itemsep=0pt]
        \item Focus on concrete results, performance improvements, or theoretical breakthroughs
        \item Include quantitative metrics when available (accuracy, efficiency, etc.)
        \item Emphasize the practical or theoretical significance of the findings
    \end{itemize}
\end{enumerate}

\vspace{0.3cm}
\textbf{Quality Guidelines:}

\begin{itemize}[leftmargin=*,itemsep=0pt]
    \item \textbf{Structure}: Exactly 3 sentences, each addressing one required element in order
    \item \textbf{Clarity}: Use precise, concrete language with specific technical terms when necessary
    \item \textbf{Specificity}: Include quantitative results when available (e.g., ``achieves 95\% accuracy'', ``reduces computation by 40\%'')
    \item \textbf{Uniqueness}: Emphasize what makes this paper distinct from related work in the same field
    \item \textbf{Distinction}: Ensure each aspect captures non-overlapping information to maximize novelty evaluation
    \item \textbf{Accessibility}: Minimize technical jargon; when used, ensure it's essential and well-defined
    \item \textbf{Completeness}: Each sentence should stand alone and be informative without requiring additional context
    \item \textbf{Novelty Focus}: Prioritize novel contributions over standard or incremental improvements
\end{itemize}

\vspace{0.3cm}
\textbf{Writing Style:}

\begin{itemize}[leftmargin=*,itemsep=0pt]
    \item Use active voice when possible for clarity and directness
    \item Start each sentence with the main point for that element to improve readability
    \item Ensure logical flow: problem $\rightarrow$ method $\rightarrow$ findings
    \item Focus on novel contributions rather than standard approaches or background information
    \item Use specific, concrete language rather than vague or generic descriptions
    \item Maintain academic tone while being accessible to researchers in related fields
\end{itemize}

\vspace{0.3cm}
\textbf{What to Avoid:}

\begin{itemize}[leftmargin=*,itemsep=0pt]
    \item Generic phrases like ``this paper presents'' or ``we propose'' without specific details
    \item Repetitive information already implied by context or common knowledge
    \item Unnecessary background information available elsewhere in the literature
    \item More or fewer than exactly 3 sentences
    \item Overly technical terms without clear necessity or explanation
    \item Sentences that don't clearly address their assigned element
    \item Overlapping content between the three aspects (each should capture distinct information)
    \item Generic descriptions that could apply to multiple papers in the same field
    \item Vague statements without concrete details or specific contributions
    \item Overly broad claims without supporting evidence or specificity
\end{itemize}

\vspace{0.3cm}
\textbf{Evaluation Criteria for Novelty:}

\begin{itemize}[leftmargin=*,itemsep=0pt]
    \item \textbf{Research Question}: Does it address a unique problem or take a novel angle on an existing problem?
    \item \textbf{Method}: Does it introduce a genuinely new technique, algorithm, or approach?
    \item \textbf{Findings}: Do the results represent significant advancement or unique insights?
\end{itemize}

\vspace{0.3cm}
\textbf{Abstract to Summarize:}

\texttt{\{abstract\}}

\vspace{0.3cm}
\textbf{Your Three-Sentence Structured Summary:}

\begin{enumerate}[leftmargin=*,itemsep=0pt]
    \item \textbf{Research Question/Problem}: [Sentence addressing the unique challenge or question that distinguishes this work from existing research in the fields]
    \item \textbf{Method/Approach}: [Sentence describing the novel technique, algorithm, or methodology that sets this work apart from previous approaches]
    \item \textbf{Key Findings/Insights/Contributions}: [Sentence stating the unique empirical results, theoretical insights, or practical contributions with specific metrics when available]
\end{enumerate}
\end{tcolorbox}

\subsection{Citation Edge Relationship Classification}
\label{sec:citation_edge_prompts}

We use Qwen3-4B-Instruct-2507 (model path: \texttt{Qwen/Qwen3-4B-Instruct-2507}) to classify citation relationships between paper pairs. The model evaluates three dimensions of citation relationships: Research Problem Relationship, Method Approach Relationship, and Key Findings Relationship, with the following configuration: max\_new\_tokens=1024 to accommodate detailed analysis, do\_sample=False for deterministic outputs, and \texttt{enable\_thinking=False} to suppress internal reasoning tokens. Below is the complete prompt used for this classification:

\begin{tcolorbox}[colback=gray!5!white, colframe=gray!75!black, breakable]
\small
\texttt{You are an expert scientific researcher analyzing citation relationships between academic papers. Your task is to evaluate WHY a citing paper (P2) references a cited paper (P1) by assessing three distinct dimensions of their relationship.}

\vspace{0.3cm}
\textbf{Input Structure:}

For both papers, you will receive:
\begin{itemize}[leftmargin=*,itemsep=0pt]
    \item \textbf{Title}
    \item \textbf{Research Question:} The core problem or question the paper addresses
    \item \textbf{Method:} The approach, techniques, or methodology used
    \item \textbf{Key Findings:} The main results, conclusions, or contributions
    \item \textbf{Citing Paper (P2):} The paper making the citation
    \item \textbf{Cited Paper (P1):} The paper being cited
\end{itemize}

\vspace{0.3cm}
\textbf{Your Task:}

Analyze the citation relationship across three dimensions and assign scores from 0.0 to 1.0 (in 0.2 increments) for each dimension. Consider what role P1 plays in P2's work.

\vspace{0.3cm}
\textbf{Dimension 1: Research Problem Relationship}

Compare how P2's research question relates to P1's research question.

\begin{itemize}[leftmargin=*,itemsep=2pt]
    \item \textbf{1.0 - Direct Extension/Continuation:} P2 explicitly builds upon, extends, or is the next step in P1's research problem. P2 directly states it is addressing limitations, gaps, or natural progressions of P1's problem. The problems form a clear lineage.
    \begin{itemize}[itemsep=0pt]
        \item Examples: ``extending X to handle Y'', ``addressing the limitation of X in scenarios Z'', ``generalizing X's problem to broader contexts''
    \end{itemize}
    
    \item \textbf{0.8 - Direct Competition (Same Problem):} P2 and P1 tackle the exact same research problem but propose different solutions. They are direct competitors addressing identical challenges. P2 likely compares its approach directly against P1 as a primary baseline.
    \begin{itemize}[itemsep=0pt]
        \item Examples: Both papers aim to solve ``few-shot image classification on ImageNet'' or ``neural machine translation for low-resource languages''
    \end{itemize}
    
    \item \textbf{0.6 - Sub-problem or Specialization:} P2 addresses a specific sub-problem, special case, or narrower aspect of P1's broader problem. P1 tackles a general problem while P2 focuses on a particular instantiation or component of it.
    \begin{itemize}[itemsep=0pt]
        \item Examples: P1 addresses ``dialogue systems'' while P2 focuses specifically on ``turn-taking in task-oriented dialogues''
    \end{itemize}
    
    \item \textbf{0.4 - Same Domain, Different Problems:} P2 and P1 work within the same research domain or application area but address distinctly different problems. They share high-level context but diverge in their specific research questions.
    \begin{itemize}[itemsep=0pt]
        \item Examples: Both in ``object detection'' but P1 focuses on ``small object detection'' while P2 addresses ``3D bounding box estimation''
    \end{itemize}
    
    \item \textbf{0.2 - Tangential/Peripheral Connection:} Very weak connection. Papers share only broad field membership or abstract conceptual links. The research questions are fundamentally different but might share very high-level themes.
    \begin{itemize}[itemsep=0pt]
        \item Examples: Both in ``computer vision'' but P1 does ``image segmentation'' while P2 does ``video captioning''
    \end{itemize}
    
    \item \textbf{0.0 - No Connection:} Research questions are unrelated. P1 is cited for purely tangential reasons (e.g., citing a standard dataset paper, a general survey, or methodological tool unrelated to the core research question).
\end{itemize}

\vspace{0.3cm}
\textbf{Dimension 2: Method Approach Relationship}

Compare how P2's methodology relates to P1's methodology.

\begin{itemize}[leftmargin=*,itemsep=2pt]
    \item \textbf{1.0 - Direct Use/Adaptation as Core Component:} P2 directly implements, adapts, extends, or modifies P1's method as a central, essential component of its approach. P1's method is foundational to P2's technical contribution. Removing P1's method would fundamentally change P2's approach.
    \begin{itemize}[itemsep=0pt]
        \item Examples: ``We adopt the X architecture from P1'', ``Building on P1's Y algorithm, we modify...'', ``Using P1's framework as our backbone...''
    \end{itemize}
    
    \item \textbf{0.8 - Primary Baseline Comparison:} P2 implements P1's method specifically to compare against it as a main baseline or competitor. P1's method represents the state-of-the-art or primary alternative that P2 aims to outperform. Substantial discussion of methodological differences.
    \begin{itemize}[itemsep=0pt]
        \item Examples: ``Compared to P1's approach which uses X, our method employs Y...''
    \end{itemize}
    
    \item \textbf{0.6 - Methodological Inspiration/Similar Framework:} P2 is conceptually inspired by P1's methodological approach or uses a similar framework/paradigm, but with significant modifications. The connection is more than superficial but P2 doesn't directly implement P1's method.
    \begin{itemize}[itemsep=0pt]
        \item Examples: ``Inspired by P1's use of attention mechanisms...'', ``Following a similar encoder-decoder structure to P1...''
    \end{itemize}
    
    \item \textbf{0.4 - Minor Component or Tool:} P2 uses a specific technique, module, or tool from P1 as one small piece of its overall method. P1's contribution is auxiliary or peripheral to P2's main methodological approach.
    \begin{itemize}[itemsep=0pt]
        \item Examples: Using P1's data augmentation technique, loss function, or evaluation protocol as one component among many
    \end{itemize}
    
    \item \textbf{0.2 - Abstract/High-Level Similarity:} Very weak methodological connection. Papers might both use the same general class of methods (e.g., ``both use transformers'' or ``both use reinforcement learning'') but in substantively different ways. The connection is at the level of paradigm rather than specific technique.
    
    \item \textbf{0.0 - No Methodological Connection:} Methods are completely unrelated. P1 might be cited for its findings, problem formulation, or dataset, but not for its methodology.
\end{itemize}

\vspace{0.3cm}
\textbf{Dimension 3: Key Findings Relationship}

Compare how P2's findings relate to and engage with P1's findings.

\begin{itemize}[leftmargin=*,itemsep=2pt]
    \item \textbf{1.0 - Foundational Premise/Dependency:} P2 explicitly relies on P1's findings as a foundational premise, assumption, or starting point for its own work. P1's findings are accepted as established facts that justify or motivate P2's approach. P2 builds its argument on P1's conclusions.
    \begin{itemize}[itemsep=0pt]
        \item Examples: ``Given P1's finding that X is crucial, we propose...'', ``P1 demonstrated that Y, therefore we investigate...''
    \end{itemize}
    
    \item \textbf{0.8 - Direct Validation/Contradiction/Discussion:} P2 directly engages with P1's specific findings—either validating, contradicting, refining, or extensively discussing them in its own experimental results section. P1's findings are a central point of discussion in P2's analysis.
    \begin{itemize}[itemsep=0pt]
        \item Examples: ``We confirm P1's finding that...'', ``Contrary to P1's results...'', ``Our experiments show P1's conclusion holds only when...''
    \end{itemize}
    
    \item \textbf{0.6 - Direct Improvement/Advancement:} P2 reports findings that directly improve upon, advance, or supersede the specific quantitative or qualitative results reported in P1. Direct performance comparison on the same metrics, datasets, or evaluation criteria.
    \begin{itemize}[itemsep=0pt]
        \item Examples: ``We improve upon P1's accuracy of X\% to Y\%'', ``While P1 achieved Z, our approach attains...''
    \end{itemize}
    
    \item \textbf{0.4 - Background Context:} P2 mentions P1's findings only as general background, related work, or literature context. P1's findings are acknowledged but not central to P2's own results or analysis. Superficial citation in related work section.
    \begin{itemize}[itemsep=0pt]
        \item Examples: ``Previous work (P1) found that...'', ``P1 reported...'', typically appearing only in introduction or related work
    \end{itemize}
    
    \item \textbf{0.2 - Minimal/Tangential Relevance:} P1's findings are barely relevant to P2. Perhaps a single finding is mentioned in passing, or the connection is extremely indirect. P1 might be cited in a list of references without specific findings being discussed.
    
    \item \textbf{0.0 - No Engagement with Findings:} P1's findings are not discussed, not relevant, or P1 is cited purely for its method, problem formulation, or as a generic reference. The findings themselves play no role in P2.
\end{itemize}

\vspace{0.3cm}
\textbf{Output Requirements:}

Step 1: Provide detailed reasoning for your scores.

Step 2: After your reasoning, provide a final JSON output.

\textbf{YOUR ENTIRE RESPONSE MUST BE A SINGLE JSON OBJECT.}

The reasoning and analysis MUST be included as fields within the JSON.

Produce a single JSON object in exactly this format:

\vspace{0.2cm}
\noindent\texttt{\{}\\
\texttt{~~"research\_problem\_score": [score between 0.0-1.0],}\\
\texttt{~~"method\_approach\_score": [score between 0.0-1.0],}\\
\texttt{~~"key\_findings\_score": [score between 0.0-1.0],}\\
\texttt{~~"reasoning": "[One concise sentence summarizing}\\
\texttt{~~~~the primary reason for the citation]",}\\
\texttt{~~"research\_problem\_analysis": "[2-4 sentences}\\
\texttt{~~~~comparing P1's research question to P2's research}\\
\texttt{~~~~question. Explicitly identify the relationship type}\\
\texttt{~~~~and justify your score with specific evidence]",}\\
\texttt{~~"method\_approach\_analysis": "[2-4 sentences}\\
\texttt{~~~~comparing P1's method to P2's method. Explicitly}\\
\texttt{~~~~identify the relationship type and justify your}\\
\texttt{~~~~score with specific evidence]",}\\
\texttt{~~"key\_findings\_analysis": "[2-4 sentences comparing}\\
\texttt{~~~~P1's key findings to P2's key findings. Explicitly}\\
\texttt{~~~~identify the relationship type and justify your}\\
\texttt{~~~~score with specific evidence]"}\\
\texttt{\}}
\vspace{0.2cm}

\textbf{IMPORTANT:} You must respond with a single JSON object and nothing else. Do not include any preamble, explanation, or markdown formatting before or after the JSON. Your entire response must be the JSON object itself, starting with '\{' and ending with '\}'.

\vspace{0.3cm}
\textbf{Paper Information:}

\vspace{0.2cm}
\noindent\texttt{<CITING PAPER (P2)>}\\
\texttt{Title: \{citing\_title\}}\\
\texttt{Research Question: \{citing\_rq\}}\\
\texttt{Method: \{citing\_method\}}\\
\texttt{Key Findings: \{citing\_findings\}}

\vspace{0.2cm}
\noindent\texttt{<CITED PAPER (P1)>}\\
\texttt{Title: \{cited\_title\}}\\
\texttt{Research Question: \{cited\_rq\}}\\
\texttt{Method: \{cited\_method\}}\\
\texttt{Key Findings: \{cited\_findings\}}

\end{tcolorbox}

\subsection{Novelty-Related Reference Extraction from Human Reviews}
\label{subsec:review_reference_extraction_prompts}

We use GPT-5.2 (via OpenAI Responses API) to extract novelty-related references from human review comments. The model analyzes reviewer feedback and categorizes cited works into three dimensions: Research Problem Overlap, Method/Approach Overlap, and Key Findings Overlap, with the following configuration: max\_output\_tokens=2048 to accommodate detailed structured output, reasoning effort set to \texttt{'medium'}  and text verbosity set to \texttt{'medium'}. Below is the complete prompt used for this extraction:

\begin{tcolorbox}[colback=gray!5!white, colframe=gray!75!black,  breakable]
\small
\texttt{You are acting as an assistant to an academic reviewer. Your task is to extract and organize references from a reviewer's comments that are relevant to evaluating the \textbf{novelty} of a submission. Novelty in this context means how the submission's contribution compares to what has been done before. A reviewer often notes prior work to highlight redundancy or overlap: perhaps the underlying \textbf{research problem} has already been studied, the \textbf{method or approach} is borrowed, or the \textbf{key findings} have already been reported. Capturing these citations systematically will help assess where the new work is or is not novel.}

\vspace{0.3cm}
\textbf{Task Instructions:}

To do this effectively, read the entire review carefully and identify every piece of prior work that the reviewer explicitly mentions or implicitly references when questioning novelty. For each identified work, decide which of the following three categories best describes why the reviewer brought it up:

\vspace{0.3cm}
\textbf{Category Definitions:}

\begin{enumerate}[leftmargin=*,itemsep=2pt]
    \item \textbf{Research problem overlap} -- the prior work poses or solves the \textit{same research problem} as the submission. For example, if the reviewer writes ``this paper revisits the same scheduling problem studied in \textit{A Unified Approach to Fair Scheduling},'' you would assign that citation to this category. Here we care about overlap in the \textit{problem statement} or \textit{objective}, regardless of how it was approached.
    
    \item \textbf{Method/approach overlap} -- the prior work uses the \textit{same algorithm, technique, or methodology} as the submission. For instance, if a reviewer says ``the authors' method is essentially the Transformer architecture introduced by \textit{Attention is All You Need},'' that reference belongs here. This category focuses on whether the procedure or algorithmic contribution is already known, even if the problem differs.
    
    \item \textbf{Key findings overlap} -- the prior work reports \textit{results or conclusions} similar to those claimed by the submission. An example would be ``\textit{Deep Residual Learning for Image Recognition} already demonstrated that skip connections improve accuracy on ImageNet,'' when the reviewer criticizes the novelty of the submission's experimental findings. This category applies when the novelty concern is about the novelty of outcomes or insights.
\end{enumerate}

\vspace{0.3cm}
\textbf{Extraction Guidelines:}

Only extract references that are actually tied to a novelty concern. Do \textbf{not} list general background citations, standard datasets (unless the review specifically notes missing comparisons), or passing mentions of common techniques that are not framed as novelty issues. The goal is to build a set of citations that the reviewer uses to question or contextualize the submission's novelty, so err on the side of excluding irrelevant or purely contextual references.

\vspace{0.3cm}
\textbf{Required Fields for Each Reference:}

\begin{itemize}[leftmargin=*,itemsep=2pt]
    \item \textbf{title} (Required): Use the most specific paper title you can. If the reviewer explicitly names the paper, reproduce the title verbatim. If only an acronym or method name is given (e.g., ``ResNet''), infer the likely full title (e.g., ``Deep Residual Learning for Image Recognition''). If you cannot confidently infer the full title, include the exact span mentioned in the review instead.
    
    \item \textbf{title\_completeness} (Required): Indicate whether the title you provided is complete (\texttt{"complete"}) or a partial/inferred title (\texttt{"incomplete"}).
    
    \item \textbf{authors} (Optional): Include authors only if the reviewer explicitly mentions them (e.g., ``Smith et al.''). Do not guess or infer authors.
    
    \item \textbf{venue} (Optional): Include the publication venue (conference, journal, etc.) only if it is explicitly mentioned by the reviewer.
    
    \item \textbf{year} (Optional): Include the publication year only if explicitly mentioned in the review.
    
    \item \textbf{context} (Required): Summarize why this reference is relevant for novelty evaluation in the chosen category---explain briefly whether it overlaps in problem, method, or findings.
    
    \item \textbf{original\_mention} (Required): Quote one or two sentences from the review that mention this reference so that another reader can verify the context.
    
    \item \textbf{novelty\_concern} (Required): Paraphrase the specific novelty issue raised, such as ``previous work addressed the same problem,'' ``uses the same method,'' or ``reports similar results.''
    
    \item \textbf{confidence} (Optional): If you wish, include a qualitative confidence level (\texttt{"high"}, \texttt{"medium"}, \texttt{"low"}) in your extraction.
\end{itemize}

\vspace{0.3cm}
\textbf{Output Format (JSON only, no other text):}

\vspace{0.2cm}
\noindent\texttt{\{}\\
\texttt{~~"research\_problem": [}\\
\texttt{~~~~\{}\\
\texttt{~~~~~~"title": "Full Paper Title Here",}\\
\texttt{~~~~~~"title\_completeness": "complete",}\\
\texttt{~~~~~~"authors": ["Author1", "Author2"],}\\
\texttt{~~~~~~"venue": "Conference/Journal",}\\
\texttt{~~~~~~"year": "2020",}\\
\texttt{~~~~~~"context": "Why this work addresses the same problem",}\\
\texttt{~~~~~~"original\_mention": "Sentence(s) from the review}\\
\texttt{~~~~~~~~mentioning the reference",}\\
\texttt{~~~~~~"novelty\_concern": "What novelty issue is raised}\\
\texttt{~~~~~~~~(problem overlap)",}\\
\texttt{~~~~~~"confidence": "high"}\\
\texttt{~~~~\}}\\
\texttt{~~],}\\
\texttt{~~"method\_approach": [}\\
\texttt{~~~~\{}\\
\texttt{~~~~~~"title": "Full Paper Title Here",}\\
\texttt{~~~~~~"title\_completeness": "complete",}\\
\texttt{~~~~~~"authors": ["Author1", "Author2"],}\\
\texttt{~~~~~~"venue": "Conference/Journal",}\\
\texttt{~~~~~~"year": "2020",}\\
\texttt{~~~~~~"context": "Why this work uses the same method or}\\
\texttt{~~~~~~~~approach",}\\
\texttt{~~~~~~"original\_mention": "Sentence(s) from the review}\\
\texttt{~~~~~~~~mentioning the reference",}\\
\texttt{~~~~~~"novelty\_concern": "What novelty issue is raised}\\
\texttt{~~~~~~~~(method overlap)",}\\
\texttt{~~~~~~"confidence": "medium"}\\
\texttt{~~~~\}}\\
\texttt{~~],}\\
\texttt{~~"key\_findings": [}\\
\texttt{~~~~\{}\\
\texttt{~~~~~~"title": "Full Paper Title Here",}\\
\texttt{~~~~~~"title\_completeness": "incomplete",}\\
\texttt{~~~~~~"authors": [],}\\
\texttt{~~~~~~"venue": null,}\\
\texttt{~~~~~~"year": null,}\\
\texttt{~~~~~~"context": "Why this work reports similar findings",}\\
\texttt{~~~~~~"original\_mention": "Sentence(s) from the review}\\
\texttt{~~~~~~~~mentioning the reference",}\\
\texttt{~~~~~~"novelty\_concern": "What novelty issue is raised}\\
\texttt{~~~~~~~~(finding overlap)",}\\
\texttt{~~~~~~"confidence": "low"}\\
\texttt{~~~~\}}\\
\texttt{~~]}\\
\texttt{\}}
\vspace{0.2cm}

If a particular category has no relevant citations, output an empty array for that category. If no novelty-relevant references are found at all, output:

\vspace{0.2cm}
\noindent\texttt{\{}\\
\texttt{~~"research\_problem": [],}\\
\texttt{~~"method\_approach": [],}\\
\texttt{~~"key\_findings": []}\\
\texttt{\}}
\vspace{0.2cm}

\textbf{Review text to analyze:}

\texttt{\{REVIEW\_TEXT\}}

\end{tcolorbox}
\subsection{Node-Level Novelty Relevance Filtering}
\label{subsec:novelty_reference_filter_prompts}

We use GPT-5.2 (via OpenAI Responses API) to filter paper references based on novelty relevance. The model compares citing and cited papers across three dimensions: Research Problem Overlap, Method/Approach Overlap, and Key Findings Overlap, with the following configuration: max\_output\_tokens=1024 for structured boolean classification, reasoning effort configurable as texttt{'medium'}), and text verbosity set to \texttt{'medium'} for balanced output. Below is the complete prompt used for this filtering:

\begin{tcolorbox}[colback=gray!5!white, colframe=gray!75!black,breakable]
\small
\texttt{You are an expert scientific researcher analyzing novelty relationships between academic papers. Your task is to evaluate whether a cited reference poses a novelty concern for a citing paper by assessing three distinct dimensions of their relationship.}

\vspace{0.3cm}
\textbf{Input Structure:}

For both papers, you will receive:
\begin{itemize}[leftmargin=*,itemsep=0pt]
    \item \textbf{Title}
    \item \textbf{Research Question:} The core problem or question the paper addresses
    \item \textbf{Method/Approach:} The approach, techniques, or methodology used
    \item \textbf{Key Findings:} The main results, conclusions, or contributions
    \item \textbf{Citing Paper (Current Paper):} The paper whose novelty we are evaluating
    \item \textbf{Cited Paper (Reference):} A paper that the current paper cites
\end{itemize}

\vspace{0.3cm}
\textbf{Your Task:}

Analyze the novelty relationship across three dimensions and assign boolean values (true/false) for each dimension. A reference poses a ``novelty concern'' if it significantly overlaps with the citing paper's contributions in any dimension, potentially undermining the citing paper's novelty claims.

\vspace{0.3cm}
\textbf{Dimension 1: Research Problem Overlap}

Does the cited paper address the same or very similar research problem/question as the citing paper?

\begin{itemize}[leftmargin=*,itemsep=2pt]
    \item \textbf{True (poses novelty concern):} The cited paper tackles the exact same research problem, or a nearly identical problem that makes the citing paper's problem formulation unoriginal. The citing paper is essentially revisiting or only slightly modifying a problem that has already been studied in the cited work.
    \begin{itemize}[itemsep=0pt]
        \item Examples: Both papers address ``federated learning with heterogeneous data distributions'' or ``neural architecture search for mobile devices''
    \end{itemize}
    
    \item \textbf{False (no novelty concern):} The cited paper addresses a different research problem, even if there are some conceptual connections or shared domain knowledge. The problems are fundamentally distinct despite potential thematic overlap.
    \begin{itemize}[itemsep=0pt]
        \item Examples: One paper studies ``image classification'' while the other studies ``object detection'', even though both are computer vision tasks
    \end{itemize}
\end{itemize}

\vspace{0.3cm}
\textbf{Dimension 2: Method/Approach Overlap}

Does the cited paper use the same or very similar methodology/approach as the citing paper?

\begin{itemize}[leftmargin=*,itemsep=2pt]
    \item \textbf{True (poses novelty concern):} The cited paper employs essentially the same algorithmic approach, technical method, or methodological framework as the citing paper. The citing paper's methodological contribution is largely anticipated or directly borrowed from the cited work.
    \begin{itemize}[itemsep=0pt]
        \item Examples: Both papers use ``contrastive learning with momentum encoders'' or ``transformer-based sequence-to-sequence models with multi-head attention''
    \end{itemize}
    
    \item \textbf{False (no novelty concern):} The cited paper uses different methods or approaches, even if there are some shared techniques or general paradigms. The methodological approaches are substantively different.
    \begin{itemize}[itemsep=0pt]
        \item Examples: One uses ``reinforcement learning'' while the other uses ``supervised learning'', despite both being machine learning approaches
    \end{itemize}
\end{itemize}

\vspace{0.3cm}
\textbf{Dimension 3: Key Findings Overlap}

Does the cited paper report similar or identical key findings/results as the citing paper?

\begin{itemize}[leftmargin=*,itemsep=2pt]
    \item \textbf{True (poses novelty concern):} The cited paper demonstrates or concludes similar key results, insights, or empirical findings as the citing paper. The citing paper's main contributions in terms of results or conclusions have already been established in the cited work.
    \begin{itemize}[itemsep=0pt]
        \item Examples: Both papers show that ``adding skip connections improves gradient flow in deep networks'' or ``self-attention mechanisms outperform RNNs on long sequences''
    \end{itemize}
    
    \item \textbf{False (no novelty concern):} The cited paper reports different findings or results, even if there are some related observations or shared datasets. The core conclusions and insights are distinct.
    \begin{itemize}[itemsep=0pt]
        \item Examples: One paper finds ``method A achieves 85\% accuracy'' while another finds ``method B achieves 87\% accuracy'' on the same dataset -- the specific findings differ
    \end{itemize}
\end{itemize}

\vspace{0.3cm}
\textbf{Output Requirements:}

A reference poses a ``novelty concern'' if ANY of the three dimensions shows significant overlap (true). If a reference shows overlap in at least one dimension, it should be considered novelty-relevant and retained for further novelty analysis.

Step 1: Provide detailed analysis for each dimension with specific evidence from the provided texts.

Step 2: After your analysis, provide a final JSON output.

\textbf{YOUR ENTIRE RESPONSE MUST BE A SINGLE JSON OBJECT.}

The analysis and comparisons MUST be included as fields within the JSON.

Produce a single JSON object in exactly this format:

\vspace{0.2cm}
\noindent\texttt{\{}\\
\texttt{~~"is\_novelty\_relevant": true/false,}\\
\texttt{~~"reasoning": "[One concise sentence summarizing whether}\\
\texttt{~~~~this reference poses a novelty concern for the citing}\\
\texttt{~~~~paper]",}\\
\texttt{~~"research\_problem\_overlap": true/false,}\\
\texttt{~~"method\_approach\_overlap": true/false,}\\
\texttt{~~"key\_findings\_overlap": true/false,}\\
\texttt{~~"research\_problem\_comparison": "[2-4 sentences}\\
\texttt{~~~~comparing the citing paper's research question to the}\\
\texttt{~~~~cited paper's research question. Explicitly state whether}\\
\texttt{~~~~there is overlap and justify your assessment with specific}\\
\texttt{~~~~evidence from the provided texts.]",}\\
\texttt{~~"method\_approach\_comparison": "[2-4 sentences comparing}\\
\texttt{~~~~the citing paper's method/approach to the cited paper's}\\
\texttt{~~~~method/approach. Explicitly state whether there is overlap}\\
\texttt{~~~~and justify your assessment with specific evidence from the}\\
\texttt{~~~~provided texts.]",}\\
\texttt{~~"key\_findings\_comparison": "[2-4 sentences comparing the}\\
\texttt{~~~~citing paper's key findings to the cited paper's key}\\
\texttt{~~~~findings. Explicitly state whether there is overlap and}\\
\texttt{~~~~justify your assessment with specific evidence from the}\\
\texttt{~~~~provided texts.]",}\\
\texttt{~~"overall\_analysis": "[2-3 sentences explaining the overall}\\
\texttt{~~~~novelty relationship between the papers and why this}\\
\texttt{~~~~reference is or isn't relevant to the citing paper's}\\
\texttt{~~~~novelty claims.]"}\\
\texttt{\}}
\vspace{0.2cm}

\textbf{IMPORTANT:} You must respond with a single JSON object and nothing else. Do not include any preamble, explanation, or markdown formatting before or after the JSON. Your entire response must be the JSON object itself, starting with '\{' and ending with '\}'.

\vspace{0.3cm}
\textbf{Paper Information:}

\vspace{0.2cm}
\noindent\texttt{<CITING PAPER (Current Paper)>}\\
\texttt{Title: \{citing\_title\}}\\
\texttt{Research Question: \{citing\_research\_question\}}\\
\texttt{Method/Approach: \{citing\_method\_approach\}}\\
\texttt{Key Findings: \{citing\_key\_findings\}}

\vspace{0.2cm}
\noindent\texttt{<CITED PAPER (Reference)>}\\
\texttt{Title: \{cited\_title\}}\\
\texttt{Research Question: \{cited\_research\_question\}}\\
\texttt{Method/Approach: \{cited\_method\_approach\}}\\
\texttt{Key Findings: \{cited\_key\_findings\}}

\end{tcolorbox}

\subsection{Transition-Level Novelty Relevance Filtering}
\label{subsec:transition_novelty_filter_prompts}

We use GPT-5.2 (via OpenAI Responses API) to filter paper references based on novelty relevance in transition spaces. The model analyzes reasoning transitions between concepts across two dimensions: Research Problem $\rightarrow$ Method/Approach (P2M) and Method/Approach $\rightarrow$ Key Findings (M2K), with the following configuration: max\_output\_tokens=4096 for detailed transition analysis, reasoning effort configurable as \texttt{'medium'}), and text verbosity set to \texttt{'medium'} for balanced output. Below is the complete prompt used for this filtering:

\begin{tcolorbox}[colback=gray!5!white, colframe=gray!75!black,  breakable]
\small
\texttt{You are an expert scientific researcher analyzing novelty relationships between academic papers in terms of their reasoning transitions. Your task is to evaluate whether a cited reference poses a novelty concern for a citing paper by assessing two distinct transition dimensions.}

\vspace{0.3cm}
\textbf{Input Structure:}

For both papers, you will receive:
\begin{itemize}[leftmargin=*,itemsep=0pt]
    \item \textbf{Research Question}
    \item \textbf{Method/Approach}
    \item \textbf{Key Findings}
    \item \textbf{Citing Paper (Current Paper):} The paper whose novelty we are evaluating
    \item \textbf{Cited Paper (Reference):} A paper that the current paper cites
\end{itemize}

\vspace{0.3cm}
\textbf{Your Task:}

Analyze the novelty relationship by examining whether the cited paper anticipates or establishes similar reasoning transitions as the citing paper. A reference poses a ``novelty concern'' if it demonstrates similar logical connections between concepts, potentially undermining the citing paper's novelty claims.

\vspace{0.3cm}
\textbf{Dimension 1: P2M Transition Similarity (Research Problem $\rightarrow$ Method/Approach)}

Based on the research problem and method/approach described, does the cited paper appear to follow a similar logical reasoning pattern when transitioning from problem to method as the citing paper?

\begin{itemize}[leftmargin=*,itemsep=2pt]
    \item \textbf{True (poses novelty concern):} The cited paper appears to connect its research problem to methodological choices through similar logical reasoning as the citing paper. The relationship between problem formulation and method selection suggests analogous intellectual approaches, even if the specific problems and methods differ.
    \begin{itemize}[itemsep=0pt]
        \item Examples: Both papers show similar problem-to-method reasoning like connecting gradient-related issues with attention mechanisms, or linking memory retention problems with replay-based solutions
    \end{itemize}
    
    \item \textbf{False (no novelty concern):} The cited paper appears to connect problem and method through fundamentally different logical reasoning. The relationship between problem formulation and method selection suggests distinct intellectual approaches, even if there might be superficial similarities in terminology.
    \begin{itemize}[itemsep=0pt]
        \item Examples: One paper connects error analysis problems with architectural innovations while another connects theoretical convergence questions with optimization algorithms
    \end{itemize}
\end{itemize}

\vspace{0.3cm}
\textbf{Dimension 2: M2K Transition Similarity (Method/Approach $\rightarrow$ Key Findings)}

Based on the method/approach and key findings described, does the cited paper appear to follow a similar logical reasoning pattern when transitioning from method application to deriving findings as the citing paper?

\begin{itemize}[leftmargin=*,itemsep=2pt]
    \item \textbf{True (poses novelty concern):} The cited paper appears to connect methodological application to key findings through similar logical reasoning as the citing paper. The relationship between method execution and result interpretation suggests analogous analytical approaches, even if the specific methods and findings differ.
    \begin{itemize}[itemsep=0pt]
        \item Examples: Both papers show similar method-to-findings reasoning like connecting attention mechanism analysis with syntactic insights, or linking representation analysis with contextual learning strategies
    \end{itemize}
    
    \item \textbf{False (no novelty concern):} The cited paper appears to connect method and findings through fundamentally different logical reasoning. The relationship between method execution and result interpretation suggests distinct analytical approaches, even if there might be superficial similarities in experimental design.
    \begin{itemize}[itemsep=0pt]
        \item Examples: One paper connects component ablation with efficiency insights while another connects mathematical analysis with theoretical understanding
    \end{itemize}
\end{itemize}

\vspace{0.3cm}
\textbf{Output Requirements:}

Return ONLY a single JSON object wrapped in a markdown code block (\texttt{```json ... ```}).

Do NOT output any additional text before or after the JSON.

Keep all string fields concise to avoid truncation.

\vspace{0.2cm}
\noindent\texttt{```json}\\
\texttt{\{}\\
\texttt{~~"is\_novelty\_relevant": true/false,}\\
\texttt{~~"reasoning": "[One concise sentence summarizing whether}\\
\texttt{~~~~this reference poses a novelty concern]",}\\
\texttt{~~"p2m\_transition\_similarity": true/false,}\\
\texttt{~~"m2k\_transition\_similarity": true/false,}\\
\texttt{~~"p2m\_transition\_analysis": "[1-2 sentences. Compare the}\\
\texttt{~~~~intellectual bridge from problem formulation to method}\\
\texttt{~~~~choice.]",}\\
\texttt{~~"m2k\_transition\_analysis": "[1-2 sentences. Compare the}\\
\texttt{~~~~intellectual bridge from method application to}\\
\texttt{~~~~findings/claims.]",}\\
\texttt{~~"overall\_analysis": "[1-3 sentences. Overall novelty}\\
\texttt{~~~~relationship summary.]"}\\
\texttt{\}}\\
\texttt{```}
\vspace{0.2cm}

\textbf{Paper Information:}

\vspace{0.2cm}
\noindent\texttt{<CITING PAPER (Current Paper)>}\\
\texttt{Research Question: \{citing\_research\_question\}}\\
\texttt{Method/Approach: \{citing\_method\_approach\}}\\
\texttt{Key Findings: \{citing\_key\_findings\}}

\vspace{0.2cm}
\noindent\texttt{<CITED PAPER (Reference)>}\\
\texttt{Research Question: \{cited\_research\_question\}}\\
\texttt{Method/Approach: \{cited\_method\_approach\}}\\
\texttt{Key Findings: \{cited\_key\_findings\}}

\end{tcolorbox}


\end{document}